\documentclass[fleqn,usenatbib]{mnras}

\usepackage{newtxtext,newtxmath}
\usepackage{comment}
\usepackage{listings}
\usepackage[T1]{fontenc}
\usepackage{xcolor, color}
\usepackage{lineno}
\DeclareRobustCommand{\VAN}[3]{#2}
\let\VANthebibliography\thebibliography
\def\thebibliography{\DeclareRobustCommand{\VAN}[3]{##3}\VANthebibliography}

\usepackage{graphicx}	% Including figure files
\usepackage{amsmath}	% Advanced maths commands
\usepackage{bm}
\usepackage{physics}
\usepackage{parskip}% http://ctan.org/pkg/parskip
\newcommand{\tempoDOS}{$\mathrm{{\scriptstyle TEMPO2}}$}

\newcommand{\temponest}{$\mathrm{{\scriptstyle TEMPONEST}}$}
\title[Anomalies in pulsar braking indices]{Stochastic and secular anomalies in pulsar braking indices}

\author[A.~F.~Vargas et al.]{\parbox{\linewidth}{\centering
Andr\'es F. Vargas$^{1,2}$\thanks{E-mail: a.vargas@unimelb.edu.au}, and
Andrew Melatos$^{1,2}$
}\\\\
$^{1}$School of Physics, University of Melbourne, Parkville, VIC 3010, Australia\\
$^{2}$OzGrav: The Australian Research Council Centre of Excellence for Gravitational-wave Discovery, University of Melbourne, Parkville, VIC 3010, Australia
}

\date{Accepted XXX. Received YYY; in original form ZZZ}

\pubyear{2022}

\begin{document}
\label{firstpage}
\pagerange{\pageref{firstpage}--\pageref{lastpage}}
\maketitle

\begin{abstract}

\noindent Stochastic and secular variations in the spin frequency $\nu$ of a rotation-powered pulsar complicate the interpretation of the measured braking index, $n$, in terms of a power-law spin-down torque $\propto \nu^{n_{\rm pl}}$. Both categories of variation can lead to anomalous braking indices, with $\vert n \vert = \vert \nu \ddot{\nu} / \dot{\nu}^2 \vert \gg 1$, where the overdot symbolizes a derivative with respect to time. Here we quantify the combined effect of stochastic and secular deviations from pure power-law spin down on measurements of $n$. Through analytic calculations, Monte Carlo simulations involving synthetic data, and modern Bayesian timing techniques, it is shown that the variance of $n$ satisfies the predictive, falsifiable formula $\langle n^{2} \rangle = (n_{\rm pl}+\dot{K}_{\rm dim})^{2}+\sigma_{\rm dim}^{2}$, where $\dot{K}_{\rm dim}$ is inversely proportional to the time-scale $\tau_K$ over which the proportionality constant of the power-law spin-down torque varies, $\sigma_{\rm dim}$ is proportional to the timing noise amplitude and inversely proportional to the square root of the total observing time, and the average is over an ensemble of random realizations of the timing noise process. The anomalous regime $\langle n^2 \rangle \gg 1$ occurs for $\dot{K}_{\rm dim} \gg 1$, $\sigma_{\rm dim} \gg 1$, or both. The sign of $n$ depends in part on the sign of $\dot{K}_{\rm dim}$, so it is possible to measure unequal numbers of positive and negative $n$ values in a large sample of pulsars. The distinguishable impact of stochastic and secular anomalies on phase residuals is quantified to prepare for extending the analysis of synthetic data to real pulsars.\\\\

\end{abstract}


%%%%%%%%%%%%%%%%%%%%%%%%%%%%%%%%%%%%%%%%%%%%%%%%%%

%%%%%%%%%%%%%%%%% BODY OF PAPER %%%%%%%%%%%%%%%%%%

\section{Introduction}
\label{Sec:Introduction}

The braking torque of a rotation-powered pulsar probes the pulsar's magnetospheric and interior physics~\citep{BlandfordRomani1988}. It can be studied by measuring the braking index,

\begin{equation}
 n = \frac{\nu \ddot{\nu}}{\dot{\nu}^2},
\label{Eq:Intro_n}
\end{equation}

where $\nu$ is the pulse frequency, and an overdot denotes a derivative with respect to time. In many plausible physical theories, the braking torque obeys a power law $\dot{\nu} \propto \nu^{n_{\rm pl}}$, whereupon pulsar timing experiments yield $n=n_{\rm pl}$ in the absence of intrinsic or instrumental stochastic fluctuations in $\nu$. Physical examples of power-law braking torques include $2 \lesssim n_{\rm pl} \leq 3$ for an extended corotating dipole magnetosphere~\citep{Melatos1997} or a force-free relativistic wind~\citep{Goldreich1970,BucciantiniThompson2006,ContopoulosSpitkovsky2006, KouTong2015}, $n_{\rm pl}=3$ for vacuum magnetic dipole radiation~\citep{GunnOstriker1969}, $n_{\rm pl}>3$ for vacuum electromagnetic radiation involving higher-order multiples~\citep{Petri2015,Petri2017,AraujoDeLorenci2024}, $n_{\rm pl}=5$ for mass quadrupole gravitational radiation~\citep{Thorne1980}, and $n_{\rm pl}=7$ for mass current gravitational radiation, e.g. r-modes \citep{PapaloizouPringle1978,Andersson1998,OwenLindblom1998}.

 In a few pulsars that are relatively free of rotational irregularities, such as timing noise or glitches, pulsar timing experiments measure $2\lesssim n \leq 3$, consistent with an electromagnetic torque~\citep{LivingstoneKaspi2007,LivingstoneKaspi2011,ArchibaldGotthelf2016}. However, in most pulsars where $n$ can be measured at all, pulsar timing experiments measure $3 \ll \vert n \vert \lesssim 10^{6}$, and $n$ is found to be negative in many objects~\citep{JohnstonGalloway1999,ChukwudeChidiOdo2016,LowerBailes2020,ParthasarathyJohnston2020}. Such high $\vert n \vert$ values are termed `anomalous'.
 
 Various phenomenological modifications of the braking law have been proposed to explain anomalous braking indices~\citep{BlandfordRomani1988}. In this paper, we focus on the interplay between two modifications: (i) the proportionality factor in $\dot{\nu} \propto \nu^{n_{\rm pl}}$ evolves secularly; and (ii) the secular torque is supplemented by a stochastic torque, which drives spin wandering and dominates $\ddot{\nu}$ over typical observational time-scales. In scenario (i), we write

\begin{equation}
    \dot{\nu} = K(t) \nu^{n_{\rm pl}}, 
    \label{Eq:secularpowerlaw}
\end{equation}

where $K(t)$ evolves on a time-scale $\tau_{K}$ much shorter than the pulsar's spin-down time-scale. Equation~(\ref{Eq:secularpowerlaw}) yields $\vert n \vert \approx \vert n_{\rm pl}  +\nu/(\dot{\nu}\tau_{K}) \vert \gg n_{\rm pl}$ with $n_{\rm pl}$ constant. Physical examples of (i) include (counter)alignment of the rotation and magnetic axes~\citep{Goldreich1970,LinkEpstein1997,Melatos2000,BarsukovPolyakova2009,JohnstonKarastergiou2017,AbolmasovBiryukov2024}, magnetic field evolution due to ohmic dissipation or Hall drift~\citep{TaurisKonar2001,GeppertRheinhardt2002,PonsVigano2012,GourgouliatosCumming2015}, precession~\citep{Melatos2000,BarsukovTsygan2010,BiryukovBeskin2012,GoglichidzeBarsukov2015,WassermanCordes2022}, and magnetospheric switching~\citep{LyneHobbs2010,StairsLyne2019,TakataWang2020,WangTakata2023}. In scenario (ii), we write

\begin{equation}
    \ddot{\nu}(t)=\ddot{\nu}_{\rm em}(t)+\zeta(t),
    \label{Eq:ddotnuscreened}
\end{equation}

where $\ddot{\nu}_{\rm em}(t)$ obeys constant-$K(t)$ power-law braking (e.g.\ with $n_{\rm pl}=3)$, and $\zeta(t)$ is a noisy driver~\citep{VargasMelatos2023}. Physical examples of (ii) include relaxation processes mediated by crust-superfluid coupling~\citep{AlparBaykal2006,GugercinogluAlpar2014,Akbal2017,Gugercinoglu2017,LowerJohnston2021} and achromatic timing noise inherent to the stellar crust or superfluid core~\citep{CordesDowns1985,AlparNandkumar1986,Jones1990,D'AlessandroMcCulloch1995,MelatosLink2014,ChukwudeChidiOdo2016}. The foregoing examples are distinct from chromatic timing noise due to propagation effects~\citep{GoncharovReardon2021},  or fluctuations confounding the measurement of $\ddot{\nu}$ for a polynomial ephemeris~\citep{ChukwudeBaiden2010,ColesHobbs2011}, which also occur in general. Population studies confirm that anomalous braking indices are correlated with glitch activity and high timing noise amplitude~\citep{ArzoumanianNice1994,JohnstonGalloway1999,UramaLink2006,LowerJohnston2021}.

In this paper, we generalize~\cite{VargasMelatos2023} to quantify the combined effect of secular $K(t)$ evolution [scenario (i) above] and a stochastic torque [scenario (ii) above] when measuring $n$ via equation~(\ref{Eq:Intro_n}). We emphasize that the treatment is phenomenological; it concentrates on analyzing the measurement process systematically but it does not follow from a first-principles theory of the braking torque. We synthetically generate pulse times of arrival (TOAs) for a statistical ensemble of pulsars obeying equation~(\ref{Eq:ddotnuscreened}), with $\nu_{\rm em}(t)$ obeying equation~(\ref{Eq:secularpowerlaw}), thereby including both scenarios (i) and (ii) simultaneously. We copy \cite{VargasMelatos2023} and use the Bayesian \temponest~timing software~\citep{ShannonCordes2010,LentatiAlexander2014,ParthasarathyJohnston2020,LowerJohnston2021} to measure $n$, via equation~(\ref{Eq:Intro_n}), from the synthetic TOAs. Every random realization of $\zeta(t)$ in equation~(\ref{Eq:ddotnuscreened}) generates a unique TOA sequence and $n$ value. In other words, there is an unavoidable physical dispersion in astronomical $n$ measurements, even in an ideal world with zero experimental errors, because there is no way to know where the actual realization of $\zeta(t)$ encountered in an astronomical timing experiment falls within the physically admissible ensemble. From these systematic numerical experiments we develop a predictive, falsifiable formula for the variance $\langle n^2 \rangle$ as a function of $n_{\rm pl}$ and the properties of $K(t)$ and $\zeta(t)$. The result generalizes the formula in~\cite{VargasMelatos2023} for $\langle n^2 \rangle$ as a function of $n_{\rm pl}$ and the properties of $\zeta(t)$, which assumes $K(t) = {\rm constant}$; see equation (15) in the latter reference.

The paper is structured as follows. Section~\ref{Sec:simulmesnandthemodel} introduces the generalized model used to generate the TOAs, i.e. equations~(\ref{Eq:secularpowerlaw}) and (\ref{Eq:ddotnuscreened}). Section~\ref{Sec:varandrmsBI} and the Appendix present the derivation and validation of the predictive, falsifiable formula for $\langle n^2 \rangle$. We validate the formula using Monte-Carlo simulations involving synthetic realizations of TOA sequences, as done previously by~\cite{VargasMelatos2023}. Section~\ref{Sec:conclusions} presents the conclusions. 

\section{Simulating braking index measurements}
\label{Sec:simulmesnandthemodel}

To quantify how secular [scenario (i)] and stochastic [scenario (ii)] torques mask $n_{\rm pl}$, we perform a numerical experiment like the one in~\cite{VargasMelatos2023}. The experiment proceeds through the following steps. (i) We generate synthetic time series of the pulse phase and its first three derivatives, according to a phenomenological Ornstein-Uhlenbeck model, which combines secular $K(t)$ evolution and a stochastic torque. These synthetic time series are converted into TOAs. (ii) The synthetic TOAs are fed into~\tempoDOS~\citep{HobbsEdwards2006}, to generate a `traditional' timing solution, as a starting point for analyzing the synthetic TOAs with~\temponest~\citep{LentatiAlexander2014}. (iii) From the~\temponest~analysis we measure $n$ and compare it with $n_{\rm pl}$. (iv) We repeat steps (i)--(iii) for an ensemble of realizations, with stochastic and secular anomalies of different strengths, to generate statistical distributions of measured $n$ values. 

This section explains the practical details involved in steps (i)--(iv) to assist the reader with reproducibility. In Section~\ref{subsec:Themodel}, we briefly introduce the phenomenological Ornstein-Uhlenbeck model used to generate the synthetic TOAs; full details can be found in~\cite{VargasMelatos2023}. Section~\ref{subsec:Theexperiment} describes the process to analyze the synthetic TOAs and measure $n$. Section~\ref{subsec:Anexample} presents a worked example involving a representative synthetic pulsar with the rotational parameters of PSR J0942--5552.  

\subsection{Secular and stochastic torques}
\label{subsec:Themodel}

A pulsar's rotational evolution is described by the rotational phase $\phi(t)$ of the crust, the rotational frequency $\nu(t)=\dot{\phi}(t)$, and the frequency derivatives $\dot{\nu}(t)$ and $\ddot{\nu}(t)$. These dynamical variables are stored in the state vector ${\bf X} = (X_1,X_2,X_3,X_4)^{\rm T} = (\phi,\nu,\dot{\nu},\ddot{\nu})^{\rm T}$, where ${\rm T}$ denotes the matrix transpose.

The phenomenological model in this paper evolves ${\bf X}$ under the action of secular and stochastic torques. For the sake of definiteness, we take the secular torque to be the standard magnetic dipole torque in equation~(\ref{Eq:secularpowerlaw}), with $n_{\rm pl}=3$~\citep{GunnOstriker1969}. Other values of $n_{\rm pl}$ are equally valid, as referenced in Section~\ref{Sec:Introduction}. In this paper, we evolve $K(t)$ according to

\begin{equation}
    K(t) = K_{2}+(K_{1}-K_{2})e^{-t/\tau_{K}},
    \label{Eq:Koft}
\end{equation}

with $K(0)=K_{1}$ and $K(\infty)=K_{2}$. Equation~(\ref{Eq:Koft}) is a phenomenological description of one possible history of $K(t)$. It is not derived from a specific physical theory of spin down and merely models a monotonic increase or decrease over a time-scale $\tau_{K}$.\footnote{Other functional forms, e.g. a sigmoid function, are equally valid as long the condition $\vert \dot{K}(t) / K(t) \vert \sim \tau_{K}^{-1}$ holds.} One plausible physical example of a process obeying equation~(\ref{Eq:Koft}) is the (counter)alignment of a pulsar's rotational and magnetic axes, during which the angle $\alpha$ between the two axes changes monotonically from $\alpha(0)=\alpha_1$ to $\alpha(\infty)=0$ [or $\alpha(\infty)=\pi/2$], with $K(t) \propto \sin^{2}\alpha(t)$ approximately~\citep{Goldreich1970,LinkEpstein1997,Melatos2000,ContopoulosSpitkovsky2006}. For many plausible mechanisms, $\tau_{K}$ is much shorter than the pulsar's spin-down time-scale, $\tau_{\rm sd} \approx \nu/\vert \dot{\nu} \vert$, with $10^{-2} \leq \tau_{K}/\tau_{\rm sd} \leq 10^{-3}$ typically for (counter)alignment as one illustrative example~\citep{Goldreich1970,Melatos2000}. In practice, the model accepts the user-selected ratios $K_{2}/K_{1}$ (which can be greater or less than unity) and $\tau_{K}/\tau_{\rm sd}$ to generate the synthetic TOAs; see Appendix~\ref{Appendix:theory}. 

The stochastic torque is modeled phenomenologically as a Langevin driver, a standard approach~\citep{MeyersMelatos2021,MeyersO'Neill2021,VargasMelatos2023,O'NeillMeyers2024}. The statistics of the Langevin driver are tailored to generate mean-reverting fluctuations in the phase residuals (or equivalently the TOAs), whose magnitude and autocorrelation time-scale are comparable to those observed in real pulsars. The mathematical form of the stochastic torque is not derived self-consistently from a physical theory; it is one of many valid options, just like the secular torque. It provides a rigorous mathematical framework for analyzing systematically the observational recipe for measuring $n$, but it is not unique. In this paper, we assume that the Langevin driver injects white noise into $\ddot{\nu}(t)$, which produces red noise in $\phi(t),\nu(t)$, and $\dot{\nu}(t)$ after integrating $\ddot{\nu}(t)$; integration acts as a low-pass filter. The justification for this choice, which includes ensuring that observables such as $\langle \ddot{\nu}(t) \ddot{\nu}(t') \rangle$ do not diverge in the limit $t \rightarrow t'$, is subtle and is discussed in detail in Section 2 and Appendix A of \cite{VargasMelatos2023}. It is easy to add additional white noise to $\nu(t)$ or $\dot{\nu}(t)$, if future observations assert the need~\citep{Cheng1987,Jones1990,GugercinogluAlpar2017,MeyersMelatos2021,MeyersO'Neill2021,AntonelliBasu2022}. The Langevin driver is calibrated by generating phase residuals which qualitatively resemble those observed in real pulsars, as explained in Section~\ref{subsec:Anexample}.

The secular and stochastic torques described above combine to yield four simultaneous stochastic differential equations of motion,

\begin{equation}
    d{\bf X}=({\bf A}{\bf X}+{\bf E})dt+{\bf \Sigma}d{\bf B}(t),
    \label{Eq:setofequations}
\end{equation}

with

\begin{equation}
    \bm A = \begin{pmatrix} 0 & 1 & 0 & 0 \\ 0 & -\gamma_{\nu} & 1 & 0\\ 0 & 0 & -\gamma_{\dot{\nu}} & 1 \\
    0 & 0 & 0 & -\gamma_{\ddot{\nu}} \end{pmatrix}, \label{Eq:Amplitudes_Matrix} 
\end{equation}

\begin{equation}
    \bm E = \begin{pmatrix} 0 \\ \gamma_{\nu} \nu_{\rm em}(t) \\ \gamma_{\dot{\nu}} \dot{\nu}_{\rm em}(t) \\ \dddot{\nu}_{\rm em}(t)+\gamma_{\ddot{\nu}} \ddot{\nu}_{\rm em}(t) \end{pmatrix}, \label{Eq:torque_vector}
\end{equation}

\noindent and

\begin{equation}
    \bm \Sigma = \text{diag}\left(0, 0, 0 ,\sigma_{\ddot{\nu}} \right). \label{Eq:Sigma_Matrix}
\end{equation}

In equations (\ref{Eq:setofequations})--(\ref{Eq:Sigma_Matrix}), the parameters $\gamma_{\nu},\gamma_{\dot{\nu}}$, and $\gamma_{\ddot{\nu}}$ are constant damping coefficients, $\nu_{\rm em}(t)$ is the solution to the secular evolution described by equations~(\ref{Eq:secularpowerlaw}) and~(\ref{Eq:Koft}), $\dot{\nu}_{\rm em}(t)$ and $\ddot{\nu}_{\rm em}(t)$ correspond to the first and second time derivatives of $\nu_{\rm em}(t)$, respectively, and $\sigma_{\ddot{\nu}}^{2}$ defines the amplitude of the Langevin driver, discussed below. We present analytic solutions for $\nu_{\rm em}(t)$ and its derivatives in Appendix~\ref{Appendix:theory}. The Wiener increment $d{\bf B}(t)$ is treated as a memoryless, white-noise process, viz.

\begin{equation}
    \langle d{\bf B}(t) \rangle = 0,
    \label{Eq:memoryless}
\end{equation}

and 

\begin{equation}
    \langle d{\bf B}_{i}(t)d{\bf B}_{j}(t') \rangle = {\delta}_{ij}\delta(t-t'),
    \label{Eq:finvarianceB}
\end{equation}

where $\langle...\rangle$ denotes the ensemble average. For simplicity, we assume there are no cross-correlations, i.e. ${\bf \Sigma}_{ij}=0$ for $i \neq j$ in equation~(\ref{Eq:Sigma_Matrix}), and the $\ddot{\nu}$ component in the diagonal is the only non-zero term, with amplitude ${\bf \Sigma}_{44}=\sigma^{2}_{\ddot{\nu}}$. Thus $\nu(t)$, and its derivatives, are differentiable quantities which can be used to calculate $n$ from equation~(\ref{Eq:Intro_n}).\footnote{For a complete justification see Section~2.1 and Appendix A1 of \cite{VargasMelatos2023}.} If the relevant physics includes fluctuations in another element of ${\bf X}$, e.g. in $\phi(t)$ due to magnetospheric fluctuations, ${\bf \Sigma}$ is easily modified, e.g. with $\Sigma_{11} \neq 0$. The reader is encouraged to experiment with a different form of ${\bf\Sigma}$, if future data demand. 

The equations of motion (\ref{Eq:Koft})--(\ref{Eq:finvarianceB}) describe an Ornstein-Uhlenbeck process, the archetype of linear, mean-reverting, random processes including Brownian motion. We emphasize, however, that equations (\ref{Eq:Koft})--(\ref{Eq:finvarianceB}) are phenomenological; they are not derived self-consistently or uniquely from a physical theory of the secular and stochastic torques referenced in Section 1. Rather, equations (\ref{Eq:Koft})--(\ref{Eq:finvarianceB}) reproduce qualitatively the impact of secular anomalies [scenario (i)] and timing noise [scenario (ii)], in the same spirit as previous analyses~\citep{MeyersMelatos2021,MeyersO'Neill2021,AntonelliBasu2022,VargasMelatos2023}. \cite{VargasMelatos2023} showed systematically through Monte-Carlo simulations that equations (\ref{Eq:setofequations})--(\ref{Eq:finvarianceB}) mimic the canonical observed behaviour of timed pulsars, i.e. $\nu(t)\approx\nu_{\rm em}(t)$ and $\dot{\nu}(t)\approx\dot{\nu}_{\rm em}(t)$ closely follow the secular evolution in equations~(\ref{Eq:secularpowerlaw}) and (\ref{Eq:Koft}) (assuming $K_{1}=K_{2}$), with root-mean-square fluctuations $\lesssim 10^{-7}$ and $10^{-1}$ per cent respectively, while $\ddot{\nu}(t)$ is noisier and often deviates from the secular trend significantly, with fluctuations of order unity. Henceforth, we refer to the phenomenological model described by equations (\ref{Eq:Koft})--(\ref{Eq:finvarianceB}) as the Brownian model. 

\subsection{Bayesian $n$ measurements}
\label{subsec:Theexperiment}

Each realization of the Brownian model yields a set of TOAs and an initial timing solution, which includes ${\bf X}(t_{0})$. In this section, we briefly describe the recipe to create TOAs from the Brownian model, and present the two-step procedure, which involves \tempoDOS~and~\temponest, used to analyze the initial timing solution for each realization to measure $n$.

The model creates synthetic TOAs for each numerical solution of equations (\ref{Eq:Koft})--(\ref{Eq:finvarianceB}) by generating a sample of times $t_{i}$ satisfying $X_{1}(t_{i})=\phi(t_{i}) \; ({\rm mod}~2\pi) = 0$, within the observation interval $0 \leq t \leq T_{\rm obs}$. Every realization of the model with a new random seed generates unique, yet statistically equivalent, TOA and ${\bf X}(t_{i})$ sequences. All synthetic TOAs are reported alongside a constant uncertainty $\Delta_{\rm TOA}$, for simplicity. The recipe to generate the $t_{i}$ samples is described in Section~2.2 of \cite{VargasMelatos2023}.\footnote{The procedure to generate synthetic data is implemented in the publicly available {\tt baboo} package at \url{http://www.github.com/meyers-academic/baboo}.}

Every realization of the Brownian model, for some choice of the model parameters $\Xi=\{\sigma^{2}_{\ddot{\nu}},K_{2}/K_{1},\tau_{K}/\tau_{\rm sd}, \gamma_{\nu},\gamma_{\dot{\nu}},\gamma_{\ddot{\nu}}\}$, comprises the initial state ${\bf X}(t_{0})$, the pulsar's right ascension (RA) and declination (DEC), and a set of $N_{\rm TOA}$ synthetic TOAs $\{t_{1},...,t_{N_{\rm TOA}}\}$ alongside their uncertainty $\Delta_{\rm TOA}$. Because we fix the values of $\gamma_{\nu},\gamma_{\dot{\nu}}$, and $\gamma_{\ddot{\nu}}$, in order to qualitatively reproduce the observed phase residuals of real pulsars, hereafter we omit these parameters (as well as $n_{\rm pl}=3$) from $\Xi$. The values used for $\gamma_{\nu},\gamma_{\dot{\nu}}$, and $\gamma_{\ddot{\nu}}$ are presented in Section~\ref{subsec:Anexample}.  The generated ephemeris containing ${\bf X}(t_{0})$, RA, and DEC and the TOA sequence are fitted using \tempoDOS~\citep{HobbsLyne2004} to create an initial estimate of the parameters $\theta=\{{\rm RA,DEC},\nu,\dot{\nu},\ddot{\nu}\}$ and their uncertainties $\Delta \theta = \{\Delta {\rm RA},\Delta {\rm DEC},\Delta \nu,\Delta \dot{\nu},\Delta \ddot{\nu}\}$.

The \tempoDOS~outputs for $\theta$ and $\Delta \theta$ are used to set the priors for \temponest, following the same procedure as in Section 2.3 of \cite{VargasMelatos2023}. \temponest~includes two phenomenological models for the red timing noise and excess white noise in the phase residuals~\citep{LentatiAlexander2014}. Red timing noise is modeled by the frequency domain power spectral density (PSD) 

\begin{equation}
    P_{\rm r}(f) = \frac{A^{2}_{\rm red}}{12\pi^{2}}\left(\frac{f}{f_{\rm yr}}\right)^{-\beta}, 
    \label{Eq:Temponest_TimingNoise}
\end{equation}

where $A_{\rm red}$ is the amplitude, $\beta$ is the spectral index, and we define $f_{\rm yr}=(1~{\rm yr})^{-1}$. \temponest~handles the excess white noise by replacing $\Delta_{\rm TOA}$ with a modified uncertainty per TOA, $\mu=({\rm EQUAD})^{2}+({\rm EFAC})\Delta_{\rm TOA}$. Here, EQUAD is the error in quadrature for the stationary excess noise, and EFAC is a fitting factor which corrects for instrumental effects~\citep{LentatiAlexander2014,LowerBailes2020,KeithNitu2023}. Table~\ref{Table_subsecII:priorsTN} summarizes the priors used throughout the paper for $\theta$, the timing noise PSD, EFAC, and EQUAD.\footnote{For a constant TOA uncertainty satisfying $\Delta_{\rm TOA}\ll 1$ (such as the one used in this paper; see Table~\ref{Table_subsecII:example_injected_values})~\temponest~typically returns $\mu \approx \Delta_{\rm TOA}$, i.e. ${\rm EQUAD}\ll 1$ and ${\rm EFAC}\approx 1$.  However, to be conservative, we choose instead the priors for EFAC and EQUAD in Table~\ref{Table_subsecII:priorsTN}, as these align with prior ranges used in real pulsar timing experiments, e.g \cite{ParthasarathyShannon2019} and ~\cite{LowerBailes2020}. The choice $\Delta_{\rm TOA}\ll 1$ helps to highlight the effect of the process noise [scenario (ii) in Section~\ref{Sec:Introduction}], relative to the effect of the measurement noise.} The center (either linear or logarithmic) of the prior range (column 2 in Table~\ref{Table_subsecII:priorsTN}) is given by the~\tempoDOS~ephemeris in the first three rows and by the midpoint of the range in column 2 in the fourth to ninth rows. 

\begin{table}
\centering
\caption{Prior ranges used by \temponest~for $\theta$ (upper half) and the timing noise parameters (lower half) used to analyze a realization of the Brownian model. The same priors are used when analysing an ensemble of realizations with fixed $\Xi$, as in Section~\ref{Sec:varandrmsBI}, except that $\ddot{\nu}$ spans $(-10^{-3} \langle \Delta\ddot{\nu} \rangle ,10^{3} \langle \Delta\ddot{\nu} \rangle)$, where $\langle \Delta \ddot{\nu} \rangle$ is averaged over the ensemble of realizations.}
\label{Table_subsecII:priorsTN}
\begin{tabular}{llll}
\hline
Parameter~[units] & Prior range & Center& Prior \\
\hline
 RA [rad] & $(-10^{-5},10^{-5})$ & \tempoDOS & Uniform \\
 DEC [rad] & $(-10^{-5},10^{-5})$ & \tempoDOS & Uniform \\
 $\nu$ [Hz] & $(-10^{-4},10^{-4})$ & \tempoDOS & Uniform \\
 $\dot{\nu}$ [Hz\;s$^{-1}$] & $(-10^{-12},0)$ & Absolute & Uniform \\
 $\ddot{\nu}$ [Hz\;s$^{-2}$] & $(-10^{-3}\Delta\ddot{\nu},10^{3}\Delta\ddot{\nu})$ & Absolute & Uniform \\
 \hline
 EFAC & $(-1,3)$ & Absolute & Uniform \\
 EQUAD [s] & $(10^{-10},10^{-2})$ & Absolute & Log-uniform \\
 $A_{\rm red}~[{\rm yr}^{3/2}]$ & $(10^{-15},10^{-5})$ & Absolute &  Log-uniform \\
 $\beta$ & $(2,10)$  & Absolute & Log-uniform\\
\hline
\end{tabular}
\end{table}

From \temponest's ephemeris, we determine $n$ through equation~(\ref{Eq:Intro_n}). Given that the fractional uncertainties for $\nu$ and $\dot{\nu}$ satisfy $\lesssim 10^{-9}$ and $\lesssim 10^{-3}$ respectively for most pulsars $(\sigma_{\ddot{\nu}}^{2} \leq 10^{-50}~{\rm Hz}^{2}\;{\rm s}^{-5})$, the formal measurement uncertainty for $n$, denoted $\Delta n$, is dominated by the uncertainty in $\ddot{\nu}$, with

\begin{equation}
    \Delta n = \frac{\nu \Delta \ddot{\nu}}{\dot{\nu}^{2}}.
    \label{Eq:deltan}
\end{equation}

From an ensemble of synthetic realizations with fixed $\Xi$, we construct probability distributions of $n$ and $\Delta n$ following the recipe above.  In a real, astronomical measurement, there is no way to know where the actual, astrophysical noise realization $d{\bf B}(t)$ falls within the ensemble of possible realizations. Consequently, there is no way to know where the real measurement of $n$ falls within the possible spread of $n$ measurements, nor how close it lies to $n_{\rm pl}$. We characterize the spread of possible $n$ measurements across the ensemble through the variance $\langle n^{2} \rangle$, where the average $\langle ... \rangle$ is taken over the ensemble of realizations at fixed $\Xi$. It is important to distinguish $\Delta n$ from $(\langle n^{2} \rangle-\langle n \rangle^{2})^{1/2}$. The former quantifies the precision of a given measurement, while the latter refers to the spread of possible outcomes, if the measurement were to be repeated (hypothetically) for another, otherwise identical pulsar. \footnote{In this paper, the quantity $\langle n^{2} \rangle$ is closely related to the dispersion across the ensemble, ${\rm DISP}(n)$, viz. equation (12) in \cite{VargasMelatos2023}. The distinction between $\langle n^{2} \rangle$ and ${\rm DISP}(n)$ is explained in Appendix~\ref{Appendix:theory}.} 

\subsection{Worked example: emulating secular and stochastic spin evolution for the representative pulsar PSR J0942--5552}
\label{subsec:Anexample}

To illustrate the measurement recipe in Section~\ref{subsec:Theexperiment}, we work step-by-step through the analysis for one representative pulsar, explaining how the synthetic data are generated and analyzed, and how $K(t)$ and  $\zeta(t)$ affect $n$. Specifically, we generate one noise realization $d{\bf B}(t)$ and solve equations (\ref{Eq:Koft})--(\ref{Eq:finvarianceB}) using the parameters in Table~\ref{Table_subsecII:example_injected_values}, which emulate PSR J0942--5552, an arbitrary but representative object satisfying $n= \nu(t_{0})\ddot{\nu}(t_{0})/\dot{\nu}(t_{0})^{2} = 4591 \gg n_{\rm pl}$~\citep{LowerBailes2020}. We set $\sigma_{\ddot{\nu}}^{2}=2\times10^{-50}~{\rm Hz}^{2}{\rm s}^{-5}$ to ensure that the synthetic phase residuals match qualitatively the real phase residuals output by~\tempoDOS, while $K_{2}/K_{1}=5\times10^{-1}$ and $\tau_{K}/\tau_{\rm sd}=1\times10^{-3}$ are characteristic theoretical values for the secular evolution of $K(t)$~\citep{Goldreich1970,LinkEpstein1997,Melatos2000}. 

\begin{table}
\centering
\caption{Injected rotational parameters and recovered values for the worked example in Section \ref{subsec:Anexample}, which demonstrates $\vert n \vert \gg n_{\rm pl}=3$ for scenarios (i) and (ii) by emulating the representative pulsar PSR J0942--5552~\protect \citep{LowerBailes2020}.  Numerals in parentheses indicate the $1\sigma$ uncertainty in the trailing digits. $K(t)$ evolves according to scenario (i), so we take $\ddot{\nu}(t_{0})=1.76\times10^{-24}~{\rm Hz\;s}^{-2}$ as opposed to $5.23\times10^{-27}~{\rm Hz\;s}^{-2}$, which assumes $K(t)={\rm constant}$ as in~\protect \cite{VargasMelatos2023}. The injected value of $\ddot{\nu}(t_{0})$ is obtained from equation~(\ref{Eq:app_initial_ddotnu}) assuming $n_{\rm pl}=3$, $\ddot{\nu}_{0}=5.23\times10^{-27}~{\rm Hz\;s}^{-2}$, and $\Xi$. The injected values yield $n=\nu(t_{0})\ddot{\nu}(t_{0})/\dot{\nu}(t_{0})^{2} = 1003 \gg n_{\rm pl}$, even before accounting for the stochastic torques. After the stochastic torques are included, the measured braking index is $n=-2.32\times10^{3}$. The parameters in the lower half of the table are used to generate the synthetic TOA time series.}
\label{Table_subsecII:example_injected_values}
\begin{tabular}{llll}
\hline
Parameter & Units & Injected value & Recovered value \\
\hline
 $\nu(t_{0})$ & $\text{Hz}$ & $1.5051430406$ & $1.5051430383(1)$\\
 $\dot{\nu}(t_{0})$ & $10^{-14}~\text{Hz s}^{-1}$ & $-5.1380792001$ & $-5.104(8)$ \\ 
 $\Ddot{\nu}(t_{0})$ & $10^{-24}~\text{Hz s}^{-2}$ & $1.76$ & $-4(2)$ \\ 
 \hline
 $\gamma_{\nu}$ & ${\rm s}^{-1}$ & $1\times10^{-13}$ & -- \\ 
 $\gamma_{\dot{\nu}}$ & ${\rm s}^{-1}$ & $1\times10^{-13}$ & --\\
 $\gamma_{\ddot{\nu}}$ & ${\rm s}^{-1}$ & $1\times10^{-6}$ & --\\
$\sigma_{\ddot{\nu}}^{2}$ & ${\rm Hz}^{2}{\rm s}^{-5}$ & $2\times10^{-50}$  & --\\ 
$K_{2}/K_{1}$ & -- & $5\times10^{-1}$ & -- \\
$\tau_{K}/\tau_{\rm sd}$ & -- & $1\times10^{-3}$ & --\\
$T_{\text{obs}}$ & $\text{days}$ & $1.314\times10^{3}$ & -- \\
$N_{\text{TOA}}$ & -- & $1.5\times10^{2}$ & -- \\ 
$\Delta_{\rm TOA}$ & $\mu \text{s}$ & $1\times10^{2}$ & -- \\ 
\hline
\end{tabular}
\end{table}

\begin{figure}
\flushleft
 \includegraphics[width=\columnwidth]{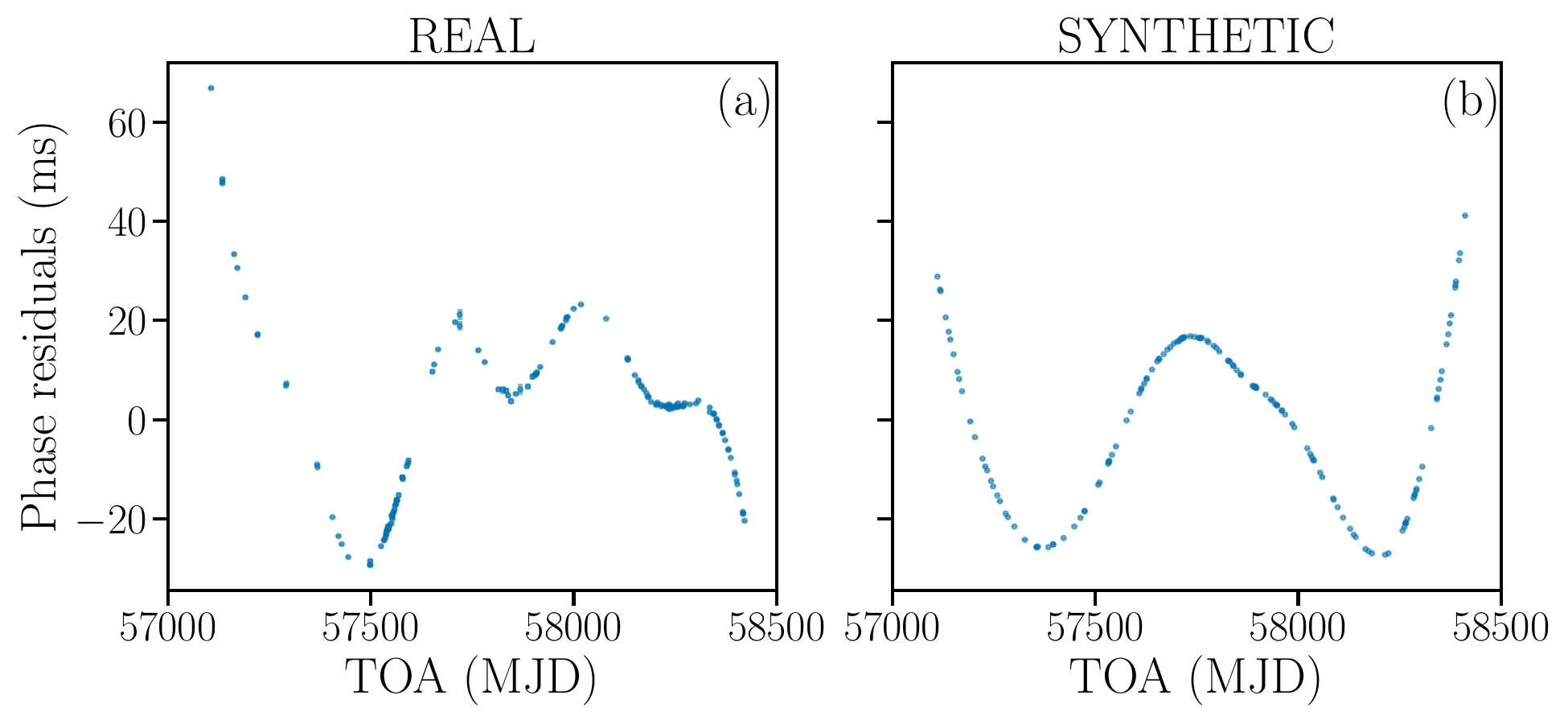}
 \includegraphics[width=\columnwidth]{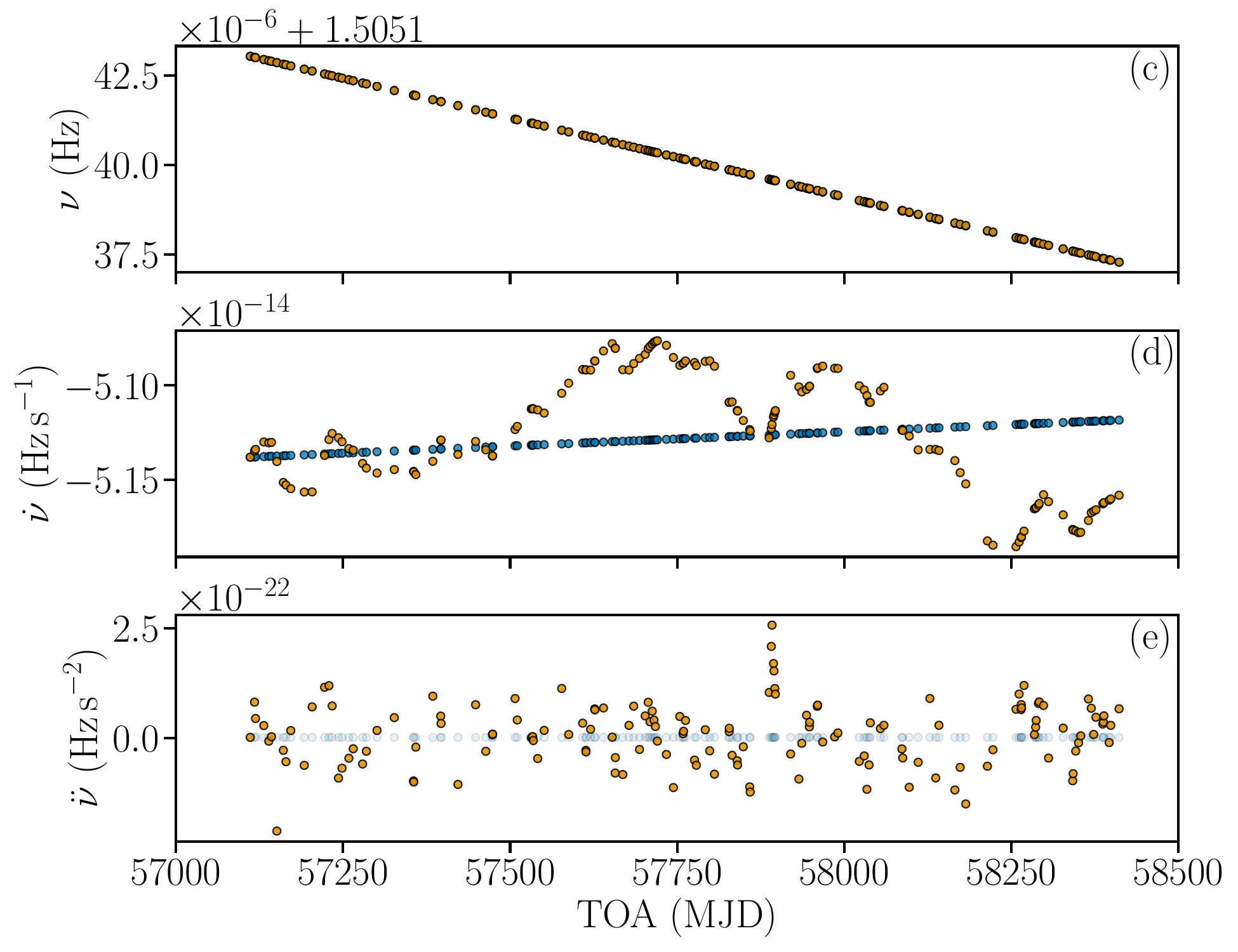}
 \caption{Actual and synthetic rotational evolution of the representative object PSR J0942$-$5552. Panels (a) and (b) show phase residuals (units: ms) versus observing epoch (units: MJD) for (a) the actual pulsar, from ~\protect \cite{LowerBailes2020}, and (b) the synthetic TOA realization, generated by solving equations (\ref{Eq:Koft})--(\ref{Eq:finvarianceB}) using the parameters in Table~\ref{Table_subsecII:example_injected_values}, respectively. The actual and synthetic phase residuals resemble each other visually. Lower panels: (c) $\nu(t)$ (orange dots) and $\nu_{\rm em}(t)$ (blue dots) (units: Hz) versus observing epoch; (d)  $\dot{\nu}(t)$ (orange dots) and $\dot{\nu}_{\rm em}(t)$ (blue dots) (units: ${\rm Hz\,s}^{-1}$) versus observing epoch; (e)  $\ddot{\nu}(t)$ (orange dots) and $\ddot{\nu}_{\rm em}(t)$ (blue dots) (units: ${\rm Hz\,s}^{-2}$) versus observing epoch. The orange and blue dots are generated by solving equations (\ref{Eq:Koft})--(\ref{Eq:finvarianceB}) for the parameters in Table~\ref{Table_subsecII:example_injected_values}, with $\sigma_{\ddot{\nu}}^{2}=0~{\rm Hz}^{2}{\rm s}^{-5}$ for the  blue dots, across all panels. One finds $\vert \ddot{\nu}(t) \vert \sim 10^{2} \vert \ddot{\nu}_{\rm em}(t) \vert$, which yields what looks approximately like a flat line for $\ddot{\nu}_{\rm em}(t)$ when compared to $\ddot{\nu}(t)$ in panel (e). The fractional fluctuations in panels (c), (d), and (e) are of order $\sim 10^{-9}$, $\sim 10^{-4}$, and $\sim 1$ respectively, in line with the actual data from PSR J0942$-$5552.}
\label{fig_subsecII:example_f2_walk}
\end{figure}

Fig.~\ref{fig_subsecII:example_f2_walk} compares visually the actual evolution of PSR J0942--5552, as observed by~\cite{LowerBailes2020}, with the synthetic evolution generated with the model in Section~\ref{subsec:Themodel}. The top panels, Figs.~\ref{fig_subsecII:example_f2_walk}(a) and~\ref{fig_subsecII:example_f2_walk}(b), present the actual and synthetic phase residuals, respectively. Both panels contain fluctuations of similar amplitude on similar time-scales. The bottom three panels, viz. Figs.~\ref{fig_subsecII:example_f2_walk}(c)--\ref{fig_subsecII:example_f2_walk}(e), display the synthetic evolution of $\nu(t), \dot{\nu}(t)$, and $\ddot{\nu}(t)$ (orange dots), which include spin wandering, and the corresponding secular trends $\nu_{\rm em}(t), \dot{\nu}_{\rm em}(t)$, and $\ddot{\nu}_{\rm em}(t)$ (blue dots), as functions of observing epoch $t$ (in units of MJD). The results reproduce the behaviour of typical timed pulsars \citep{ParthasarathyJohnston2020,LowerBailes2020}: (i) $\nu(t)$ spins down approximately linearly, with $\nu(t)\approx\nu_{\rm em}(t)$ and $T_{\rm obs} \ll \tau_{\rm sd}$; (ii) $\vert \dot{\nu}(t)\vert \propto K(t)$ decreases on the time-scale $\tau_{\rm sd}$, with small fluctuations of fractional amplitude $\sim10^{-4}$ around $\dot{\nu}_{\rm em}(t)$; and (iii) $\vert \ddot{\nu}(t) \vert \gg \vert \ddot{\nu}_{\rm em}(t) \vert $ wanders appreciably, with fluctuations of fractional amplitude $\sim 1$ exceeding the secular trend caused by $K(t)$, with $| \ddot{\nu}(t) | \sim 10^{2} |\ddot{\nu}_{\rm em}(t) |$. The injected value for $\ddot{\nu}(t_{0})=1.76\times10^{-24}~{\rm Hz\;s}^{-2}$ follows from the secular evolution of $K(t)$, i.e. from equation~(\ref{Eq:app_initial_ddotnu}) assuming $\ddot{\nu}_{0}=5.23\times10^{-27}~{\rm Hz\;s}^{-2}$, $n_{\rm pl}=\nu(t_{0})\ddot{\nu}_{0}/\dot{\nu}(t_{0})^{2}=3$, and $\Xi$. For this example, the secular evolution of $K(t)$ on its own ($\sigma^{2}_{\ddot{\nu}}=0$) yields $n=\nu(t_{0})\ddot{\nu}(t_{0})/\dot{\nu}(t_{0})^{2} = 1003 \gg n_{\rm pl}$.

Having verified in Fig.~\ref{fig_subsecII:example_f2_walk} that the synthetic phase residuals and rotational evolution are broadly realistic, we analyze the synthetic data to measure $n$. The synthetic TOAs in Fig.~\ref{fig_subsecII:example_f2_walk}(b) and ${\bf X}(t_{0})$ are fed to \tempoDOS~to obtain $\theta$ and $\Delta \theta$. \tempoDOS's estimates of $\theta$ and $\Delta \theta$ set the priors for \temponest~following the recipe in Section~\ref{subsec:Theexperiment}. Table~\ref{Table:tempo2_params_table} summarizes the estimates of $\theta, \Delta \theta$, and their associated \temponest~prior ranges. \temponest~ estimates $\theta,\Delta \theta, {\rm log} A_{\rm red}$, and $\beta$ from the synthetic TOAs and the priors in Table~\ref{Table_subsecII:priorsTN}. The inferred noise parameters are ${\rm log} A_{\rm red}=-9.17\pm0.08$ and $\beta=6.04\pm0.39$, in agreement with the reported values $\log A_{\rm red}=-9.03\pm0.2$, and $\beta=5.88^{+1.6}_{-1.1}$ in~\cite{LowerBailes2020}. The rotational parameters recovered by \temponest~are recorded in the right-hand column in Table~\ref{Table_subsecII:example_injected_values}. The fractional differences between the injected and recovered values are $\sim10^{-9}$,$\sim10^{3}$, and $\sim4$ for $\nu,\dot{\nu}$, and $\ddot{\nu}$ respectively. We combine the recovered parameters through equations~(\ref{Eq:Intro_n}) and (\ref{Eq:deltan}) to obtain $n\pm\Delta n=-2318.8\pm866.2$. The synthetic $n\pm\Delta n$ range does not overlap with the astrophysical value $n=4591^{+3.1}_{-3.5}$ quoted by \cite{LowerBailes2020}. This is expected: there is no reason why the random noise realization plotted in Fig.~\ref{fig_subsecII:example_f2_walk} should match exactly the astrophysical noise realization, even though we tune $\sigma_{\ddot{\nu}}$, so that the statistics of the synthetic and astrophysical noises are broadly similar in their orders of magnitude. Both the synthetic and astrophysical $n$ values are dispersed randomly, with variance $\langle n^2 \rangle^{1/2} \sim 10^{3}$. Any other synthetic noise realization, drawn from the same random ensemble, produces $-\langle n^2 \rangle^{1/2} \lesssim n \lesssim \langle n^2 \rangle^{1/2}$, including possibly (by pure chance) $\ddot{\nu} \approx 8\times10^{-24}~{\rm Hz\;s}^{-2}$ and hence $n\approx 4591$.

\begin{table}
\caption{Prior ranges set from \tempoDOS~estimates of $\theta$ and $\Delta \theta$ for the worked example in Section~\ref{subsec:Anexample}. $\Delta \theta$ is given by the values in parentheses in the third column, i.e. the $1\sigma$ uncertainty in the trailing digits.}
\flushleft
\begin{tabular}{p{1cm}p{1.5cm}p{2cm}p{3cm}}
\hline
Parameter & Units & \tempoDOS~estimate & Prior range \\
\hline
RA & ${\rm rad}$ & $2.540548(7)$ & $(2.54053,2.54055)$ \\
DEC & ${\rm rad}$ &  $-0.975319(5)$ &$(-0.97533,-0.97531)$ \\
 $\nu$ & ${\rm Hz}$ & $1.5051430332(6)$ &  $(1.5050,1.5052)$ \\
$\dot{\nu}$ & $ 10^{-14}~{\rm Hz\,s}^{-1}$ & $-5.108(3)$ & $(-1\times10^{2},0)$\\
$\ddot{\nu}$ & $10^{-26}~{\rm Hz\,s}^{-2}$ & $38(47)$ & $(-4.7\times10^{4},4.7\times10^{4})$\\
\hline
\end{tabular}
\label{Table:tempo2_params_table}
\end{table} 

The previous example serves to illustrate two important points. First, equations~(\ref{Eq:secularpowerlaw}) and~(\ref{Eq:Koft}) with $n_{\rm pl}=3$ generate anomalous braking indices even for $\sigma^{2}_{\ddot{\nu}}=0$, as recognized widely in the literature~\citep{Goldreich1970, BlandfordRomani1988, LinkEpstein1997,Melatos2000, TaurisKonar2001,PonsVigano2012,JohnstonKarastergiou2017,WassermanCordes2022}. The secular evolution of $K(t)$ implies [see equation~(\ref{Eq:app_initial_ddotnu})] $\ddot{\nu}(t_{0}) \propto \ddot{\nu}_{0}(\tau_{\rm sd}/\tau_{\rm K})\gg \ddot{\nu}_{0}$ and hence $n=\nu(t_{0})\ddot{\nu}(t_{0})/\dot{\nu}(t_{0})^{2} \gg n_{\rm pl}$. Second, astrophysically motivated timing noise amplitudes like $\sigma^{2}_{\ddot{\nu}}=2\times10^{-50}~{\rm Hz}^{2}\;{\rm s}^{-5}$ [see Figs.~\ref{fig_subsecII:example_f2_walk}(a) and~\ref{fig_subsecII:example_f2_walk}(b)] mask the secular evolution of $\ddot{\nu}_{\rm em}(t)$ and also contribute to the anomalous braking index through scenario (ii). For instance, the fractional difference between the injected $\ddot{\nu}(t_{0})$ and the measured $\ddot{\nu}(t_{0})$ is $\approx 3.3$. In the absence of timing noise, i.e. equations (\ref{Eq:Koft})--(\ref{Eq:finvarianceB}) with $\sigma^{2}_{\ddot{\nu}}=0$ [blue dots in Figs.~\ref{fig_subsecII:example_f2_walk}(c) -- \ref{fig_subsecII:example_f2_walk}(e)], \tempoDOS~estimates for $\nu$, $\dot{\nu}$, and $\ddot{\nu}$ yield fractional differences with the injected parameters of $\sim10^{-12},\sim10^{-6}$, and $10^{-3}$, respectively. The two previous points illustrate how secular and stochastic torques produce anomalous braking indices individually and in tandem. We postpone disentangling the individual contribution of the secular and stochastic torques, especially with reference to the phase residuals, to Section~\ref{subsec:phase_res_sto_sec_torques}.

\section{Variance of measured braking indices}
\label{Sec:varandrmsBI}

In this section, we use the model in Section~\ref{Sec:simulmesnandthemodel} to quantify the dispersion in $n$ measurements caused by the interplay between the secular and stochastic torques. Specifically, we seek a condition on $\Xi$, i.e. $K_{2}/K_{1}$, $\tau_{K}/\tau_{\rm sd}$, and $\sigma^{2}_{\ddot{\nu}}$, that yields $\langle n^{2} \rangle \gg n_{\rm pl}^{2}$, so that one is likely to measure $\vert n \vert \gg n_{\rm pl}$, when a single noise realization is drawn from a random ensemble, as in real pulsar timing experiments. To achieve this, we present in Section~\ref{subsec:nrms_and_MC} a predictive, falsifiable formula for $\langle n^{2} \rangle$, as a function of $K_{2}/K_{1}$, $\tau_{K}/\tau_{\rm sd}$, and $\sigma^{2}_{\ddot{\nu}}$, and check its validity with Monte-Carlo simulations using synthetic data created with different combinations of $\Xi$. The formula for $\langle n^{2} \rangle$ is derived rigorously in Appendix~\ref{Appendix:theory}. Section~\ref{subsec:phase_res_sto_sec_torques} examines the phase residuals produced by the limiting cases involving only secular ($\sigma^{2}_{\ddot{\nu}}=0$), or stochastic ($K_{1}=K_{2}$) torques, to set the scene, followed by the realistic case where both secular and stochastic torques act in tandem. Section~\ref{subsec:k_sign_and_populations} calculates systematically the distribution of measured $n$ values returned by~\temponest~for different secular or stochastic torque strengths. Section~\ref{subsec:nrms_vs_Ared_beta} connects the results in Section~\ref{subsec:k_sign_and_populations} to the standard parametrization of~\temponest's PSD phase residuals, namely $A_{\rm red}$ and $\beta$ in equation~(\ref{Eq:Temponest_TimingNoise}).

\subsection{$\langle n^{2} \rangle$ versus $\Xi$}
\label{subsec:nrms_and_MC}

A central question in studies of anomalous braking indices is whether one can predict the variance $\langle n^{2} \rangle$ of braking index measurements theoretically, even though of course it is impossible to predict $n$ theoretically for the specific random realization of the stochastic torque in an individual pulsar. Encouragingly, the answer to this question is yes. The proof is presented in Appendix~\ref{Appendix:theory}, where we solve the Brownian model equations (\ref{Eq:Koft})--(\ref{Eq:finvarianceB}) analytically in the regime relevant to pulsar timing experiments, with $\nu(t)\approx \nu_{\rm em}(t)$, $\dot{\nu}(t) \approx \dot{\nu}_{\rm em}(t)$, and $\vert \ddot{\nu}(t) \vert \gg \vert \ddot{\nu}_{\rm em}(t) \vert$. The result reads 

\begin{equation}
    \langle n^{2} \rangle =  \Big(n_{\rm pl}+\dot{K}_{\rm dim}\Big)^{2}+\sigma^{2}_{\rm dim},
    \label{Eq:theory_nrms}
\end{equation}

with

\begin{equation}
    \dot{K}_{\rm dim}=(n_{\rm pl}-1)\Big(1-\frac{K_{2}}{K_{1}}\Big)\frac{\tau_{\rm sd}}{\tau_{K}},
    \label{Eq:K_dim}
\end{equation}

and 

\begin{equation}
    \sigma_{\rm dim} = \frac{\sigma_{\ddot{\nu}}\nu(t_{0})}{\gamma_{\ddot{\nu}}\dot{\nu}(t_{0})^{2}T_{\rm obs}^{1/2}}.
    \label{Eq:Sigma_dim}
\end{equation}

Equation~(\ref{Eq:theory_nrms}) implies, that anomalous measurements of $n$ occur, when either the dimensionless secular coefficient, $\dot{K}_{\rm dim}$, or the dimensionless stochastic coefficient, $\sigma_{\rm dim}$, or both are large. Contours of constant $\langle n^{2} \rangle^{1/2}$ are semicircles centered at $\dot{K}_{\rm dim}=-n_{\rm pl}$ and $\sigma_{\rm dim}=0$. From equation~(\ref{Eq:K_dim}), $\dot{K}_{\rm dim}>0$ requires $K_{2}/K_{1} < 1$ while $\dot{K}_{\rm dim}<0$ requires $K_{2}/K_{1} > 1$. In the absence of secular torques, with $\dot{K}_{\rm dim}=0$ and $K_{2}/K_{1}=1$, equation~(\ref{Eq:theory_nrms}) in this paper reduces to equation~(14) in \cite{VargasMelatos2023}.

 Equation~(\ref{Eq:theory_nrms}) is a practical, falsifiable formula which, once validated, can be applied to observational studies of any rotation-powered pulsar. To validate equation~(\ref{Eq:theory_nrms}) we perform the following experiment. We select five equally log-spaced values of $\langle n^{2} \rangle^{1/2}$ within the anomalous range $3\times10^{2} \leq \langle n^{2} \rangle^{1/2} \leq 6\times10^{3}$, namely $\langle n^{2} \rangle^{1/2}_{\rm inj}=[300,634.05,1340.05,2832.18,5985.79]$. A braking index measurement in this range is likely statistically to return $\vert n \vert \gg n_{\rm pl}$. The upper limit of the range implies, via equations~(\ref{Eq:theory_nrms}) and (\ref{Eq:Sigma_dim}), a maximum noise amplitude of $\sigma^{2}_{\ddot{\nu}}=10^{-50}~{\rm Hz}^{2}{\rm s}^{-5}$, which creates astrophysical residuals that qualitatively match the real ones (see Section~\ref{subsec:Anexample}). For each value of $\langle n^{2} \rangle^{1/2}_{\rm inj}$, we draw 100 random pairs $(\dot{K}_{\rm dim},\sigma_{\rm dim})$ satisfying equation~(\ref{Eq:theory_nrms}) by writing $\sigma_{\rm dim}=\langle n^{2} \rangle^{1/2}_{\rm inj}\sin\psi$ and $\dot{K}_{\rm dim}=\langle n^{2} \rangle^{1/2}_{\rm inj}\cos\psi-n_{\rm pl}$, where $\psi$ is a random angle drawn uniformly from the interval $[2.25^{\circ},87.75^{\circ}]$ for $\dot{K}_{\rm dim}>0$ with $K_{2}/K_{1}=0.5$. Similarly, for $\dot{K}_{\rm dim}<0$, we set $K_{2}/K_{1}=1.5$ and select $\psi$ uniformly from the interval $[92.25^{\circ},177.75^{\circ}]$.\footnote{ We avoid $\psi \approx 0,90^{\circ},180^{\circ}$, as they produce $\sigma^{2}_{\ddot{\nu}}$ values, via equation~(\ref{Eq:Sigma_dim}), which generate synthetic phase residuals that differ qualitatively from the real phase residuals [Fig.~\ref{fig_subsecII:example_f2_walk}(a)] for $\langle n^{2} \rangle^{1/2}_{\rm inj}=5985.79$. For $\psi\approx 0^{\circ}$ and $180^{\circ}$ the synthetic phase residuals are orders of magnitude smaller than the real ones. The opposite is true for $\psi\approx 90^{\circ}$; see Section~\ref{subsec:phase_res_sto_sec_torques}.} Each $(\dot{K}_{\rm dim},\sigma_{\rm dim})$ pair is converted to a pair $\tau_{K}/\tau_{\rm sd}$ and $\sigma^{2}_{\ddot{\nu}}$, via equations~(\ref{Eq:K_dim})--(\ref{Eq:Sigma_dim}),  and combined with $\gamma_{\nu},\gamma_{\dot{\nu}},$ and $\gamma_{\ddot{\nu}}$ in Table~\ref{Table_subsecII:example_injected_values} to form $\Xi$. We find that $\sigma^{2}_{\ddot{\nu}}$ and $\tau_{K}/\tau_{\rm sd}$ span the ranges $1\times10^{-55} \leq \sigma^{2}_{\ddot{\nu}}/(1~{\rm Hz}^{2}\;{\rm s}^{-5}) \leq 3\times10^{-53}$ and $3\times10^{-3} \leq \tau_{K}/\tau_{\rm sd} \leq 1\times10^{-1}$ for $\langle n^{2} \rangle^{1/2}_{\rm inj}=300$, and $3\times10^{-53} \leq \sigma^{2}_{\ddot{\nu}}/(1~{\rm Hz}^{2}\;{\rm s}^{-5}) \leq 1\times10^{-50}$ and $1\times10^{-4} \leq \tau_{K}/\tau_{\rm sd} \leq 4\times10^{-3}$ for $\langle n^{2} \rangle^{1/2}_{\rm inj}=5985.79$. We generate 100 synthetic TOA sequences for the 100~$\Xi$ values and compare the root mean square of the measured $n$ values obtained from \temponest~output, $\langle n^{2} \rangle^{1/2}_{\rm meas}$, with the injected $\langle n^{2} \rangle^{1/2}_{\rm inj}$. We repeat this procedure for all five values of $\langle n^{2} \rangle^{1/2}_{\rm inj}$ and $\dot{K}_{\rm dim}>0$ and $\dot{K}_{\rm dim}< 0$. 

\begin{figure}
\flushleft
 \includegraphics[width=\columnwidth]{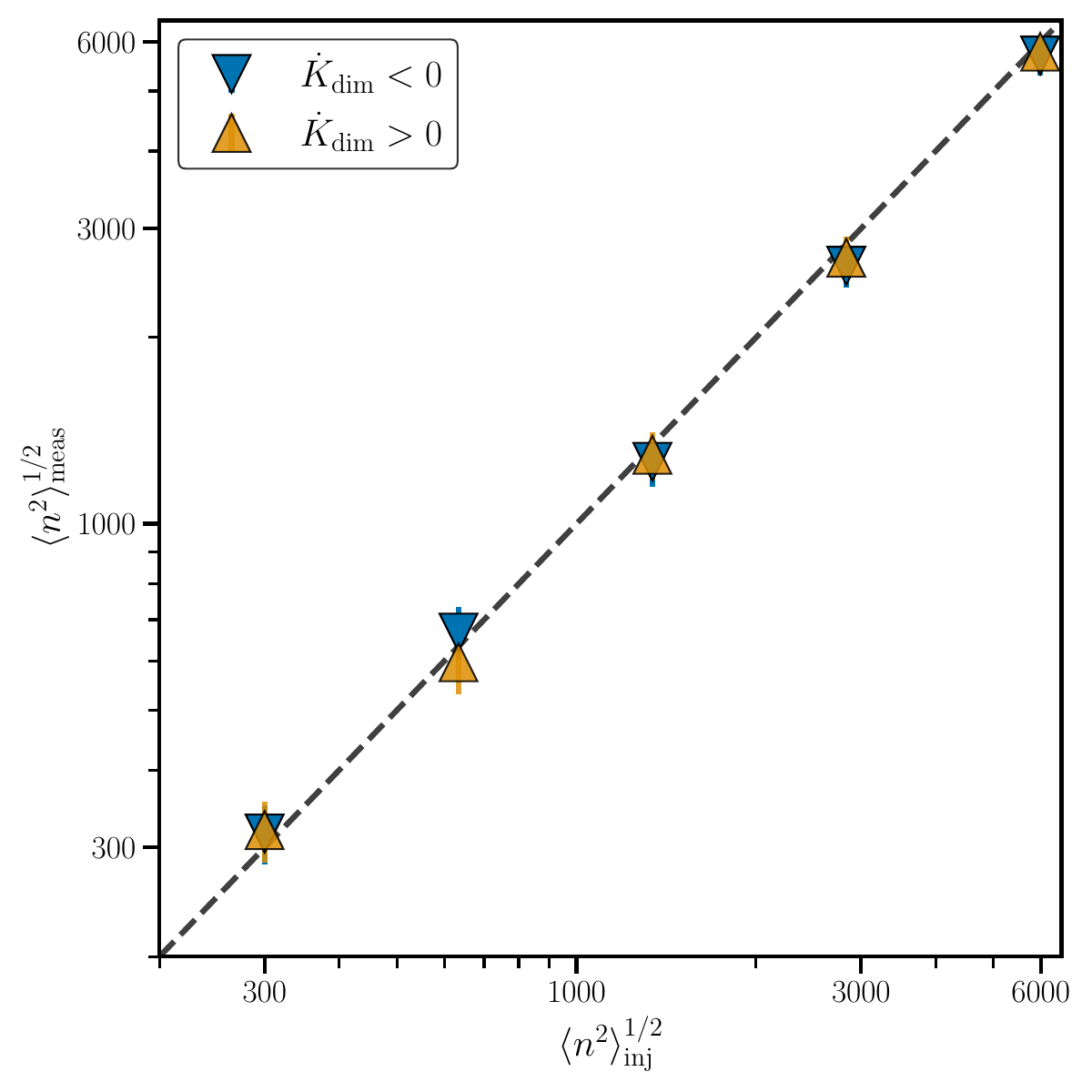}
 \caption{Validation of equation~(\ref{Eq:theory_nrms}): injected $\langle n^{2} \rangle^{1/2}_{\rm inj}$ (horizontal axis) versus recovered $\langle n^{2} \rangle_{\rm meas}^{1/2}$ (vertical axis) for synthetic data generated with $\dot{K}_{\rm dim}<0$ (blue triangles) and $\dot{K}_{\rm dim}>0$ (orange triangles) [refer to equation~(\ref{Eq:theory_nrms})]. The dotted line denotes $\langle n^{2} \rangle_{\rm meas}^{1/2}=\langle n^{2} \rangle_{\rm inj}^{1/2}$. Vertical blue and orange error bars represent the uncertainty in $\langle n^{2} \rangle^{1/2}_{\rm meas}$, calculated by propagating the uncertainty in $\Delta n$ [refer to equation~(\ref{Eq:deltan})] per realization, for $\dot{K}_{\rm dim}<0$ and $\dot{K}_{\rm dim}>0$, respectively. The maximum fractional error between $\langle n^{2} \rangle_{\rm meas}^{1/2}$ and $\langle n^{2} \rangle^{1/2}_{\rm inj}$ is $7.8\%$ for $\dot{K}_{\rm dim}< 0$ and $\langle n^{2} \rangle_{\rm inj}^{1/2}=2832.18$. The mean fractional error, averaging over all $\dot{K}_{\rm dim} >0$ and $\dot{K}_{\rm dim} <0$ realizations, is less than $6\%$.}
\label{fig_subsecIII:n_rms_mes_vs_n_rms_inj}
\end{figure}

Fig.~\ref{fig_subsecIII:n_rms_mes_vs_n_rms_inj} displays $\langle n^{2} \rangle_{\rm inj}^{1/2}$ (horizontal axis) versus $\langle n^{2} \rangle_{\rm meas}^{1/2}$ (vertical axis) for $\dot{K}_{\rm dim} >0$ (orange triangles) and $\dot{K}_{\rm dim}<0$ (blue triangles). The dotted line represents $\langle n^{2} \rangle_{\rm meas}^{1/2}=\langle n^{2} \rangle_{\rm inj}^{1/2}$. The short, vertical, orange and blue error bars represent the uncertainty in $\langle n^{2} \rangle_{\rm meas}^{1/2}$, calculated by propagating the uncertainty $\Delta n$. The validation is successful. The maximum fractional error between $\langle n^{2} \rangle_{\rm meas}^{1/2}$ and $\langle n^{2} \rangle_{\rm inj}^{1/2}$ is $7.8\%$ for $\dot{K}_{\rm dim}< 0$ and $\langle n^{2} \rangle_{\rm inj}^{1/2}=2832.18$. On average, the fractional errors from all $\dot{K}_{\rm dim}>0$ and $\dot{K}_{\rm dim}<0$ realizations are $5\%$ and $6\%$, respectively. 

The agreement between $\langle n^{2} \rangle^{1/2}_{\rm meas}$ and $\langle n^{2} \rangle^{1/2}_{\rm inj}$ confirms the validity of the synthetic measurement strategy using \temponest, and verifies equation~(\ref{Eq:theory_nrms}). With these results, it is possible to use (\ref{Eq:theory_nrms}) in real astronomical situations to predict or interpret the measured braking index (which is likely to satisfy $n\sim \langle n^{2} \rangle^{1/2}$ in a statistical sense) by tuning $\sigma_{\ddot{\nu}}/\gamma_{\ddot{\nu}}$ to qualitatively match the measured phase residuals; typically one assumes the fiducial value $\gamma_{\ddot{\nu}} = 10^{-6}{\rm s}^{-1}$~\citep{PriceLink2012,MeyersMelatos2021,MeyersO'Neill2021, VargasMelatos2023,O'NeillMeyers2024}. In practice, one possible strategy involves estimating $\langle n^{2} \rangle$ by tuning $\sigma_{\ddot{\nu}}$ to match the phase residuals, with $\dot{K}_{\rm dim}=0$ and $\sigma_{\rm dim} \neq 0$, for 100 (say) synthetic data realizations with a fixed value for $n_{\rm pl}$. Equation~(\ref{Eq:theory_nrms}) can be solved to obtain other possible $(\dot{K}_{\rm dim},\sigma_{\rm dim})$ pairs which are consistent with the estimated $\langle n^{2} \rangle$. From these pairs, equations (\ref{Eq:K_dim}) and (\ref{Eq:Sigma_dim}) yield approximate constraints for the astrophysically interesting quantities $(1-K_{2}/K_{1})\tau_{K}^{-1}$ and $\sigma_{\ddot{\nu}}$, noting that $\nu(t_{0}),\dot{\nu}(t_{0})$, and $T_{\rm obs}$ are observables. The process of estimating $\langle n^{2} \rangle$ can be repeated for synthetic realizations with different values of $n_{\rm pl}$.

\subsection{Phase residuals and the trade-off between stochastic and secular torques}
\label{subsec:phase_res_sto_sec_torques}

To illustrate the interplay between $\dot{K}_{\rm dim}$ and $\sigma_{\rm dim}$, we return briefly to the representative example of pulsar PSR J0942--5552~\citep{LowerBailes2020} from Section~\ref{subsec:Anexample}. The parameters listed in Table~\ref{Table_subsecII:example_injected_values} generate synthetic phase residuals which qualitatively resemble the real ones, with $\langle n^{2} \rangle^{1/2}=7632$, or equivalently $\dot{K}_{\rm dim}=1000$ and $\sigma_{\rm dim}=7566$. Let us consider three distinct scenarios where $\langle n^{2} \rangle^{1/2}=7632$ is caused by (a) a secular torque only ($\sigma_{\rm dim}=0$), (b) the equal combination $\dot{K}_{\rm dim}=\sigma_{\rm dim}$, and (c) a stochastic torque only ($\dot{K}_{\rm dim}=0$). Equations~(\ref{Eq:theory_nrms}) and (\ref{Eq:K_dim}) for scenario (a) imply $\tau_{\rm sd}/\tau_{K} =7632$, i.e. $\tau_{K}=61~{\rm yr}$ assuming $\tau_{\rm sd} = \nu(t_{0})/2\vert \dot{\nu}(t_{0})\vert$, and $\sigma_{\ddot{\nu}}^{2}=0$. Scenario (b) has $\dot{K}_{\rm dim}=\sigma_{\rm dim}=(\langle n^{2} \rangle / 2)^{1/2}$, whereupon equations~(\ref{Eq:theory_nrms})--(\ref{Eq:Sigma_dim}) imply $\tau_{\rm sd}/\tau_{K}=5397$ ($\tau_{K}=86~{\rm yr}$), and $\sigma^{2}_{\ddot{\nu}}=1\times10^{-50}~{\rm Hz}^{2}\;{\rm s}^{-5}$. Scenario~(c) is obtained by setting $\sigma_{\rm dim}=\langle n^{2} \rangle^{1/2}$, whereupon equations~(\ref{Eq:theory_nrms}) and (\ref{Eq:Sigma_dim}) imply $\sigma^{2}_{\ddot{\nu}}=2\times10^{-50}~{\rm Hz}^{2}\;{\rm s}^{-5}$. We set $K_{2}/K_{1}=1$ in order to achieve $\dot{K}_{\rm dim}=0$ [see equation~(\ref{Eq:K_dim})]. For each scenario we create a synthetic data realization and repeat the recipe described in Section~\ref{subsec:Anexample} to measure $n$ with \temponest.

\begin{figure}
\flushleft
 \includegraphics[width=\columnwidth]{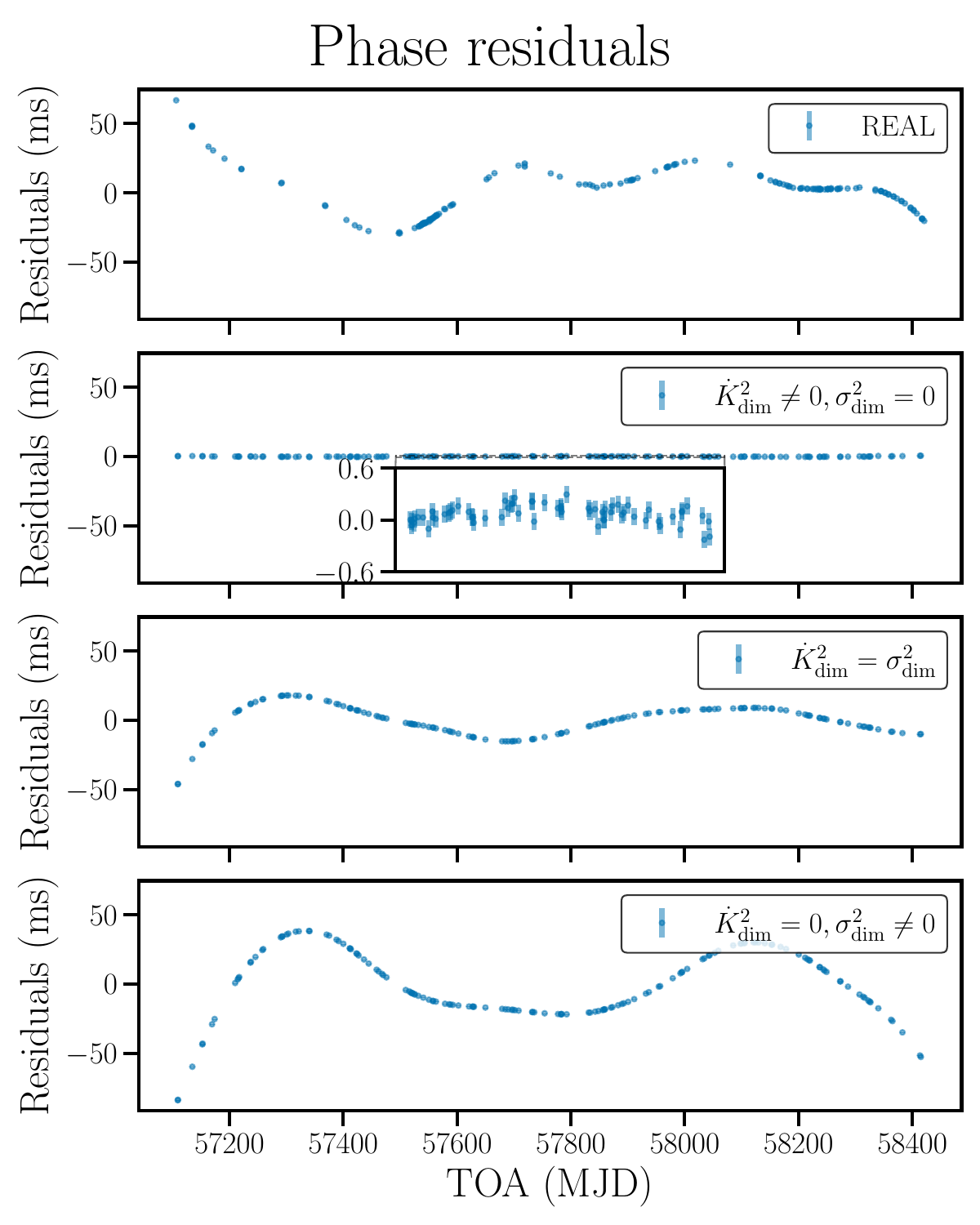}
 \caption{Phase residuals (vertical axis; units ${\rm ms}$) versus observing epoch (horizontal axis; units: MJD) for PSR J0942$-$5552 taken from \protect \cite{LowerBailes2020} (top panel), and solutions to the Brownian model equations (\ref{Eq:Koft})--(\ref{Eq:finvarianceB}) for three distinct scenarios (lower panels): secular torque only ($\dot{K}^{2}_{\rm dim} \neq 0, \sigma^{2}_{\rm dim}=0$; second panel), equal contributions from secular and stochastic torques ($\dot{K}_{\rm dim}=\sigma_{\rm dim}$; third panel), and stochastic torque only ($\dot{K}^{2}_{\rm dim} = 0, \sigma^{2}_{\rm dim}\neq 0$; fourth panel). The $\dot{K}^{2}_{\rm dim} \neq 0, \sigma^{2}_{\rm dim}=0$ scenario generates phase residuals two orders of magnitude smaller than those measured astronomically. For clarity, we plot a magnified portion of the phase residuals for this scenario (inset, second panel). The other two scenarios generate phase residuals which are visually consistent with those actually observed by~\protect \cite{LowerBailes2020}. Synthetic measurements in panels two to four return $n=7635, 3677$, and $-15102$, respectively, while the astronomical braking index is $n=4591^{+3.1}_{-3.5}$~\protect \citep{LowerBailes2020}. }
\label{fig_subsecIII:comparison_residuals_Kdim_vs_S}
\end{figure}

Fig.~\ref{fig_subsecIII:comparison_residuals_Kdim_vs_S} presents the phase residuals (vertical axis; units: ms) versus observing epoch (units: MJD) for the actual pulsar~\citep{LowerBailes2020} (first panel), as well as scenarios (a) ($\dot{K}_{\rm dim} \neq 0, \sigma_{\rm dim}=0$; second panel), (b) ($\dot{K}_{\rm dim} = \sigma_{\rm dim}$; third panel), and (c) ($\dot{K}_{\rm dim} = 0, \sigma_{\rm dim} \neq 0$; last panel). Visually the phase residuals for scenarios (b) and (c) are comparable to the actual measured ones, while for scenario (a) the synthetic phase residuals are two orders of magnitude smaller ($\leq 0.6\, {\rm ms}$; see magnified inset in second panel). The small phase residuals for scenario~(a) arise, because the secular torque produces $\ddot{\nu}(t_{0}) \approx \ddot{\nu}_{0}(\tau_{\rm sd}/\tau_{K})\approx 1.34\times10^{-23}~{\rm Hz}\;{\rm s}^{-2}$ via equation~(\ref{Eq:app_initial_ddotnu}). The \tempoDOS~fit for $\ddot{\nu}(t_{0})$ returns a fractional  of $2.8\%$ when compared to $\ddot{\nu}(t_{0})$, and an uncertainty of $\Delta \ddot{\nu}=4.8\times10^{-27}~{\rm Hz}\;{\rm s}^{-2}$.\footnote{Given the small uncertainty for scenario (a), we set \temponest's~prior range for $\ddot{\nu}$ to $(-10^{4}\Delta \ddot{\nu},10^{4}\Delta \ddot{\nu})$ as opposed to the prior listed in Table~\ref{Table_subsecII:priorsTN}.}  

\temponest~measures $n=7635\pm24,~3677\pm993$, and $-15102\pm1332$ for scenarios (a), (b), and (c), respectively. The recovered value of $n$ for scenario (a) is consistent with the injected value $n=7634$, obtained from $\nu(t_{0})$ and $\dot{\nu}(t_{0})$ in Table~\ref{Table_subsecII:example_injected_values}, and $\ddot{\nu}(t_{0})=1.34\times10^{-23}~{\rm Hz}\;{\rm s}^{-2}$. The fractional error is $0.013\%$. For scenario (a) the range $n\pm\Delta n$ excludes $n_{\rm pl}$. In scenario (b), the range $n\pm\Delta n$ overlaps with the astrophysical value $n=4591^{+3.1}_{-3.5}$ from \cite{LowerBailes2020} without including $n_{\rm pl}$. This is expected. \cite{VargasMelatos2023} confirmed systematically what is widely recognized in the literature, that the formal uncertainty on $n$, obtained via equation~(\ref{Eq:deltan}), underestimates the dispersion in $n$ values associated with different random realizations of timing noise.  

The illustrative exercise above exemplifies a central result of the paper. Secular torques can be tuned to reproduce a desired anomalous braking index but they may or may not --- by themselves --- produce phase residuals consistent qualitatively with those measured astrophysically. Scenario (a) produces a measured $n$ of the same order of magnitude as PSR J0942--5552, but the synthetic phase residuals are two orders of magnitude smaller than the real ones. In contrast, secular and stochastic torques acting in tandem [scenario~(b)] reproduce the observed phase residuals qualitatively, as does a stochastic torque acting alone [scenario (c)].

Taken at face value, it may seem surprising that the action of the secular torque [scenario (a)] is able to produce anomalous braking indices, even though it does not produce phase residuals consistent with those measured for real data, c.f. the first and second panels in Fig.~\ref{fig_subsecIII:comparison_residuals_Kdim_vs_S}. This is explained as follows. In the absence of a stochastic torque, the phase evolution of the Brownian model [(\ref{Eq:Koft})--(\ref{Eq:finvarianceB})] is given by $\phi(t) = \int_{t_{0}}^{t} dt'\,\nu_{\rm em}(t')$, where $\nu_{\rm em}(t')$ is the deterministic solution to equation~(\ref{Eq:secularpowerlaw}) and~(\ref{Eq:Koft}) for $n_{\rm pl}=3$ [see equation~(\ref{Eq:appnuSEC})]. As the time-scale of the secular torque variation is longer than $T_{\rm obs}$, with $\tau_{K} \approx 15 T_{\rm obs}$, equation~(\ref{Eq:Koft}) is approximately given by $K(t)\approx K_{1}$ for $0 \leq t \leq T_{\rm obs}$. The latter approximation implies that the phase evolution for scenario (a) is well described in terms of the polynomial ephemeris $\phi(t) = \int_{t_{0}}^{t} dt'\,\nu_{\rm em}(t') \approx \phi_{0}+\nu(t_{0})(t-t_{0})+\dot{\nu}(t_{0})(t-t_{0})^{2}/2+\ddot{\nu}(t_{0})(t-t_{0})^{3}/6$, where --- due to the secular torque --- the cubic coefficient is $\ddot{\nu}(t_{0}) \approx \ddot{\nu}_{0}(\tau_{\rm sd}/\tau_{K})\gg \ddot{\nu}_{0}$ [see equation~(\ref{Eq:app_initial_ddotnu})], implying $n= \nu(t_{0})\ddot{\nu}(t_{0})/\dot{\nu}(t_{0})^{2} \gg n_{\rm pl}$. In tandem, the $\ddot{\nu}(t_{0})(t-t_{0})^{3}/6$ term contributes $\approx 3.27$ cycles to $\phi(t)$. Not accounting for it yields cubic residuals of size $\leq 100~{\rm ms}$. Given that $\phi(t)$ can be approximated as a polynomial ephemeris for $t 
\leq T_{\rm obs}$, and that $\ddot{\nu}(t_{0})(t-t_{0})^{3}/6$ generates cubic residuals with $\leq 100~{\rm ms}$, both \tempoDOS~and~\temponest~are able to estimate $\nu(t_{0})$ and higher derivatives with high accuracy [e.g. $2.8\%$ fractional error for $\ddot{\nu}(t_{0})$], generating phase residuals consistent with the error in fitting $\ddot{\nu}(t_{0})$, i.e. $\leq 0.6~{\rm ms}$ (see magnified inset in Fig.~\ref{fig_subsecIII:comparison_residuals_Kdim_vs_S}).

\begin{figure}
\flushleft
 \includegraphics[width=\columnwidth]{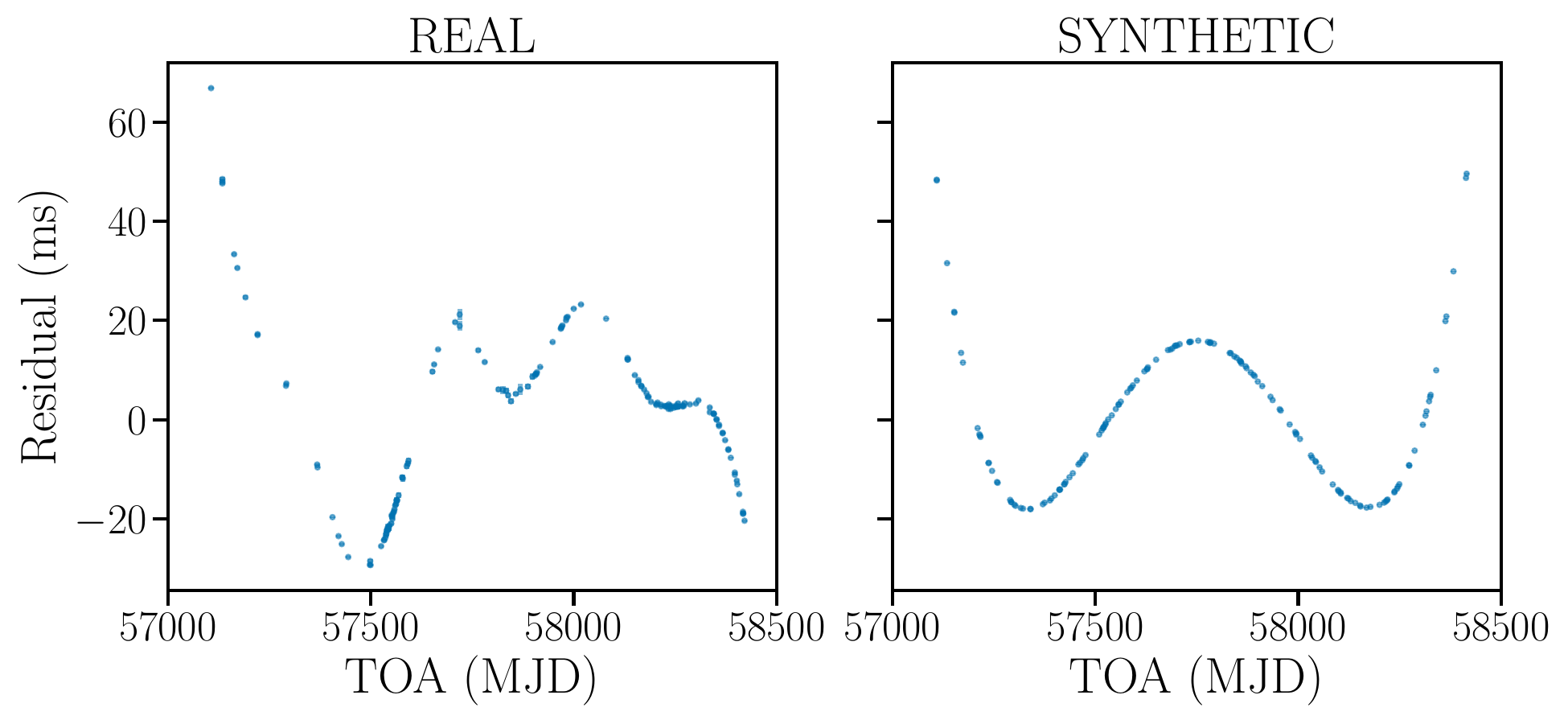}
 \caption{Effect of artificially increasing $\tau_{K}/\tau_{\rm sd}$: phase residuals (vertical axis; units ms) of PSR J0942$-$5552 \protect \citep{LowerBailes2020} (left panel) and a synthetic data realization with $\dot{K}^{2}_{\rm dim} \neq 0$ and $\sigma^{2}_{\rm dim}=0$ (right panel), versus observing epoch (horizontal axis; units: MJD). Unlike in Fig.~\ref{fig_subsecIII:comparison_residuals_Kdim_vs_S}, we adjust $\tau_{K}/\tau_{\rm sd}$, so that the left and right panels resemble each other qualitatively. The measured $n$ for this realization is $n=6.7\times10^{4}$, one order of magnitude higher than the value $n=4591^{+3.1}_{-3.5}$~measured astronomically~\protect \citep{LowerBailes2020}.}
\label{fig_subsecIII:forced_residuals}
\end{figure}

One may argue that scenario (a) fails to reproduce the observed phase residuals, because $\tau_{\rm sd}/\tau_K$ is chosen incorrectly. It turns out that this is false. If we tune $\tau_{\rm sd}/\tau_K$ to match the phase residuals, the resulting $n$ is much higher than observed. We test this by generating a new realization of scenario~(a), with $K_{2}/K_{1}=0.5$ and $\tau_{\rm sd}/\tau_{K}=9\times10^{4}$ ($\tau_{K}=5~{\rm yr}$). The results appear in Fig.~\ref{fig_subsecIII:forced_residuals}. The synthetic phase residuals (right panel) resemble visually the real ones observed in PSR J0942$-$5552 (left panel). Yet~\temponest~output yields $n=6.7\times10^{4}$ for this realization, well above the astrophysical value $n=4591^{+3.1}_{-3.5}$~\citep{LowerBailes2020}. In contrast with the previous realization of scenario (a), we observe two distinctions. First, the time-scale of the secular torque is comparable to the observation time in the repeated experiment, with $\tau_{K} \approx T_{\rm obs}$, so one cannot approximate $\phi(t)$ accurately by a polynomial ephemeris for $t \leq T_{\rm obs}$. Second, via equation~(\ref{Eq:app_initial_ddotnu}), the repeated experiment has $\tau_{\rm sd}/\tau_{K}=9\times10^{4}$ and hence $\ddot{\nu}(t_{0}) \approx 1.58\times10^{-22}~{\rm Hz}\;{\rm s}^{-2}$, as opposed to the previous realization, which has $\tau_{\rm sd}/\tau_{K}=5397$ and $\ddot{\nu}(t_{0}) \approx 1.34\times10^{-23}~{\rm Hz}\;{\rm s}^{-2}$. The first property explains the phase residuals in Fig.~\ref{fig_subsecIII:forced_residuals}, while the second property explains why the measurement $n=6.7\times10^{4}$ is one order of magnitude higher than $n=4591^{+3.1}_{-3.5}$ [astrophysical value~\citep{LowerBailes2020}] and $n=7635\pm24$ [previous realization of scenario (a)].

\subsection{Probability density function (PDF) of $n$ measurements}
\label{subsec:k_sign_and_populations}

In this subsection, we present the PDF of measured $n$ values obtained from \temponest~output for the $\langle n^{2} \rangle^{1/2}_{\rm inj}$ values tested in Section~\ref{subsec:nrms_and_MC}, for $\dot{K}_{\rm dim} > 0$ and $\dot{K}_{\rm dim} < 0$. The goals are to understand if the measurements and their error bars, $n\pm\Delta n$, bracket $n_{\rm pl}$, and if there are any population-wide differences between the $\dot{K}_{\rm dim} > 0$ and $\dot{K}_{\rm dim} < 0$ ensembles at fixed $\langle n^{2} \rangle^{1/2}_{\rm inj}$. 

\begin{figure}
\flushleft
 \includegraphics[width=\columnwidth]{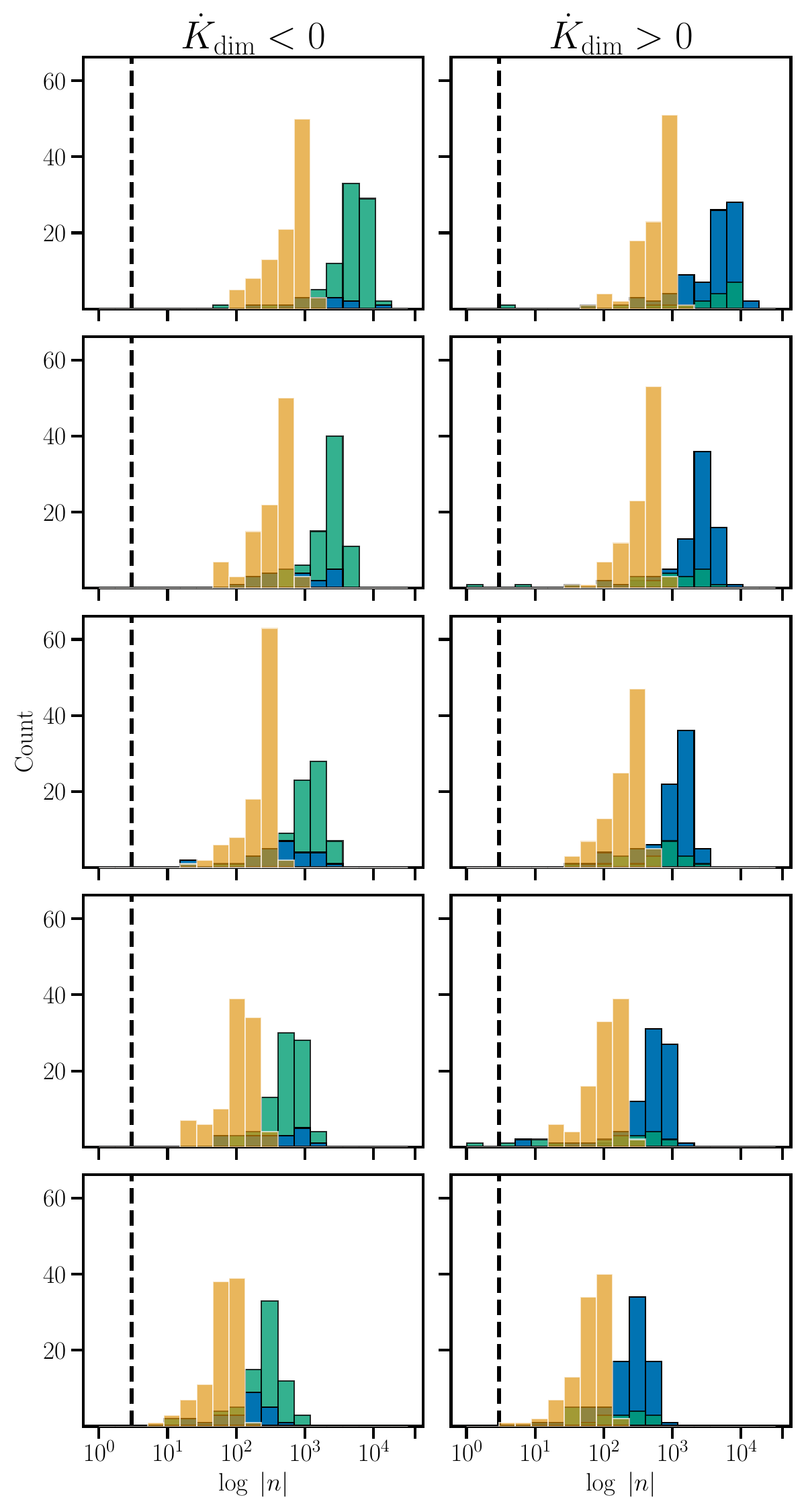}
 \caption{PDF of $\log_{10} \vert n \vert$ measurements (blue histograms for $n>0$, cyan histograms for $n<0$) and their formal uncertainty $\Delta n$ (orange histograms), calculated from \temponest~output and equation~(\ref{Eq:deltan}), for the $\langle n^{2} \rangle_{\rm inj}^{1/2}$ range tested in Section~\ref{subsec:nrms_and_MC} viz. $3\times10^{2} \leq \langle n^{2} \rangle_{\rm inj}^{1/2} \leq 6\times10^{3}$. The left and right columns record 100 trials each with $\dot{K}_{\rm dim} < 0$ and $\dot{K}_{\rm dim} > 0$, respectively. The black dotted line marks the injection $n_{\rm pl}=3$. In the bottom left panel, the average is $\langle n \rangle=-178.6$ with only 16 out of 100 measurements satisfying $n-\Delta n \leq n_{\rm pl} \leq n+\Delta n$. In the bottom right panel the average is $\langle n \rangle=200.6$ and only 15 out of 100 measurements satisfy $n-\Delta n \leq n_{\rm pl} \leq n+\Delta n$. In the top left and right panels the averages are $\langle n \rangle=-4.3\times10^{3}$ with 11 out of 100 measurements satisfying $n-\Delta n \leq n_{\rm pl} \leq n+\Delta n$, and $\langle n \rangle=3.3\times10^{3}$ with 7 out of 100 measurements satisfying $n-\Delta n \leq n_{\rm pl} \leq n+\Delta n$, respectively. The bin width in every histogram is $0.24$ dex.}
\label{fig_subsecIII:hists_kneg_kplus}
\end{figure}

Fig.~\ref{fig_subsecIII:hists_kneg_kplus} shows the PDFs of $n >0$ (blue histograms) and $n<0$ (cyan histograms) measurements, as well as the formal uncertainty $\Delta n$ obtained from \temponest~output and equation~(\ref{Eq:deltan}) (orange histograms). In the figure, $\langle n^{2} \rangle_{\rm inj}^{1/2}$ increases from the bottom panel ($\langle n^{2} \rangle^{1/2}_{\rm inj}=300$) to the top panel ($\langle n^{2} \rangle^{1/2}_{\rm inj}=5985.79$) for $\dot{K}_{\rm dim} < 0$ (left column) and $\dot{K}_{\rm dim} > 0$ (right column). The histograms are plotted on a logarithmic scale, ${\rm log}_{10} \vert n \vert$, for visual clarity.  Firstly, we observe that the orange histograms are narrower than the blue and cyan histograms for $\langle n^{2} \rangle^{1/2}_{\rm inj} \geq 300$, as found also by~\cite{VargasMelatos2023}. The blue and cyan histograms in the bottom panel span $\approx1.6$ decades, while the orange histograms span $\approx1.5$ decades. This increases to $\approx2.6$ decades for the blue and cyan histograms for the top panel, while the orange histograms span $\approx1.2$ decades. As foreshadowed in Section~\ref{subsec:phase_res_sto_sec_torques}, Fig.~\ref{fig_subsecIII:hists_kneg_kplus} implies that the formal uncertainty on $n$ [equation~(\ref{Eq:deltan})] underestimates the true dispersion in $n$ values associated with the ensemble of random realizations, one of which is the (unknown) actual realization encountered in an astronomical observation. Secondly, we observe that fewer of the 100 synthetic measurements place $n_{\rm pl}$ inside the measurement error bars $n\pm \Delta n$, as $\langle n^2 \rangle_{\rm inj}^{1/2}$ increases. In the bottom panel, only 15 and 16 realizations out of 100 satisfy $n-\Delta n \leq n_{\rm pl} \leq n+\Delta n$ for $\dot{K}_{\rm dim} < 0$ and $\dot{K}_{\rm dim} > 0$, respectively. In the top panel, only 7 and 11 realizations out of 100 satisfy $n-\Delta n \leq n_{\rm pl} \leq n+\Delta n$. By contrast, in all panels at least 60 out of 100 trials satisfy $n-\langle n^{2} \rangle^{1/2}_{\rm inj} \leq n_{\rm pl} \leq n+\langle n^{2} \rangle^{1/2}_{\rm inj}$.

Fig.~\ref{fig_subsecIII:hists_kneg_kplus} also reveals an asymmetry between the sign of the measured $n$ in the $\dot{K}_{\rm dim} < 0$ and $\dot{K}_{\rm dim} > 0$ ensembles. In all the $\dot{K}_{\rm dim} < 0$ panels, the blue and cyan histograms contain $\approx 20$ and $\approx 80$ values of $n$, respectively. The opposite is true in all the $\dot{K}_{\rm dim} > 0$ panels, with the blue and cyan histograms containing $\approx 80$ and $\approx 20$ values of $n$. This is a consequence of the interplay between the secular and stochastic torques. The secular torque sets the sign of $\ddot{\nu}(t_{0})$, via equation~(\ref{Eq:app_initial_ddotnu}). Hence one has $\ddot{\nu}(t_{0})  \propto \ddot{\nu}_{0}(\tau_{\rm sd}/2\tau_{K})$ for $\dot{K}_{\rm dim} > 0$ ($K_{2}/K_{1}=0.5$), while one has $\ddot{\nu}(t_{0})  \propto -\ddot{\nu}_{0}(\tau_{\rm sd}/2\tau_{K})$ for $\dot{K}_{\rm dim} < 0$ ($K_{2}/K_{1}=1.5$). The stochastic torque causes $\ddot{\nu}(t)$ to exhibit fluctuations of fractional amplitude $\gtrsim 1$, since for all trials we have $\sigma^{2}_{\ddot{\nu}} \geq 10^{-55}~{\rm Hz}^{2}\;{\rm s}^{-5}$. The fluctuations do not favor either sign and mask partially the secular evolution $\ddot{\nu}_{\rm em}(t)$. \footnote{Timing noise amplitudes such as $\sigma^{2}_{\ddot{\nu}} \geq 10^{-55}~{\rm Hz}^{2}\;{\rm s}^{-5}$ fall within region~(ii) of Fig. 3 in \cite{VargasMelatos2023}, where $\langle n^{2} \rangle$ (or equivalently $\langle \ddot{\nu}^{2} \rangle$) grows linearly with $\sigma^{2}_{\ddot{\nu}}$.} Coincidentally, the example in Section~\ref{subsec:Anexample} illustrates one instance where the injected and recovered values of $\ddot{\nu}(t_{0})$ have opposite signs due to the stochastic torque.

\subsection{\temponest~phase residual spectrum}
\label{subsec:nrms_vs_Ared_beta}

Equation (\ref{Eq:theory_nrms}) involves the timing noise amplitude through the Ornstein-Uhlenbeck parameters $\sigma_{\ddot{\nu}}$ and $\gamma_{\dot{\nu}}$. In the pulsar timing literature, it is common practice instead to parametrize the timing noise process in terms of the observables $A_{\rm red}$ and $\beta$ from equation~(\ref{Eq:Temponest_TimingNoise}), which describe the amplitude and shape respectively of \temponest's phase residual PSD $P_{\rm r}(f)$~\citep{LentatiAlexander2014,LentatiShannon2016,GoncharovReardon2021}. In this section, we explain how to relate $A_{\rm red}$ and $\beta$ to $\sigma_{\ddot{\nu}}$ and $\gamma_{\ddot{\nu}}$ and hence $\dot{K}_{\rm dim}$ and $\sigma_{\rm dim}$ in equation~(\ref{Eq:theory_nrms}).

\begin{figure*}
%\flushleft
 \includegraphics[width=0.8\textwidth]{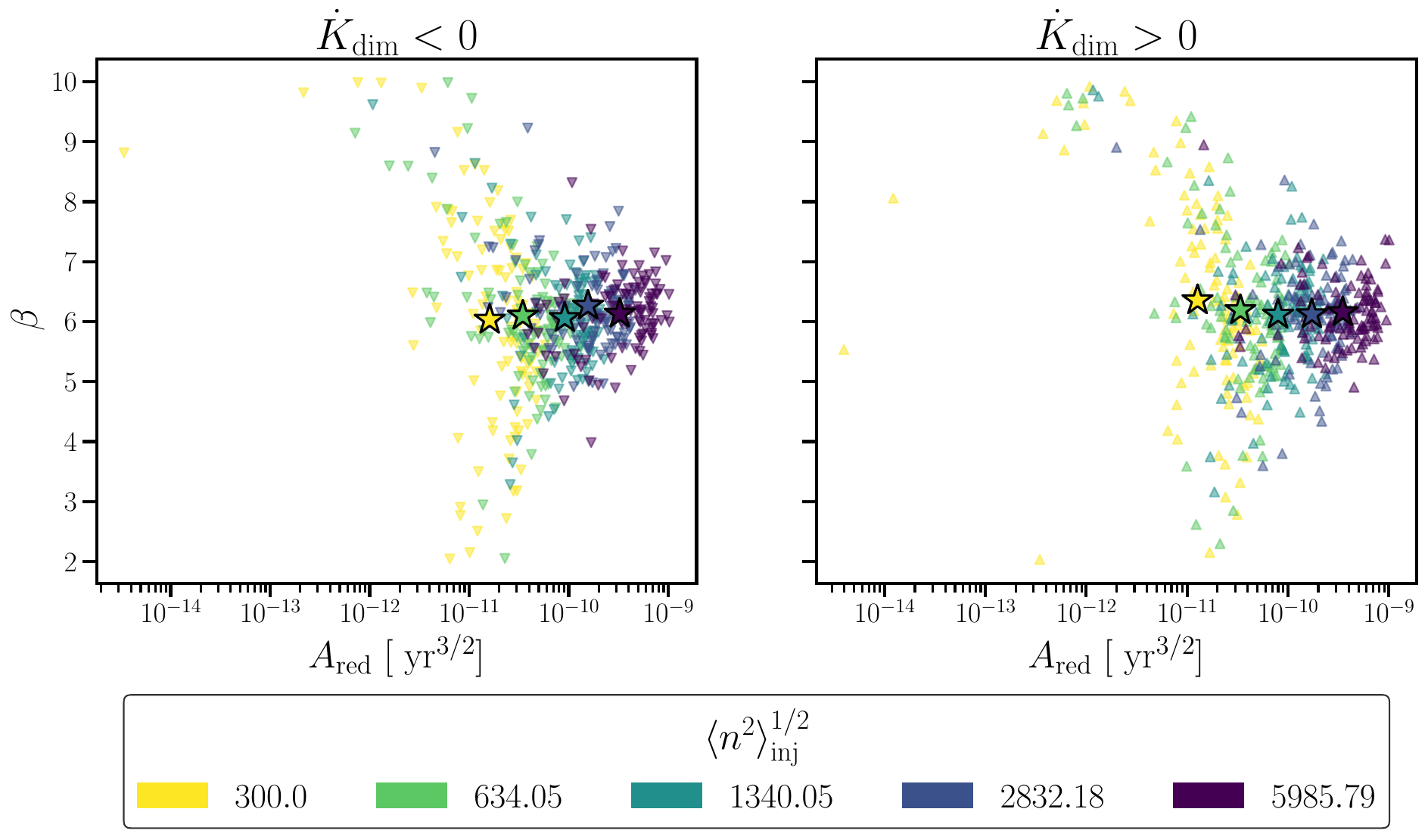}
 \caption{Phase residual PSD parameters $A_{\rm red}$ (units: ${\rm yr}^{3/2}$) and $\beta$ from~(\ref{Eq:Temponest_TimingNoise}) inferred by \temponest~as a function of injected $\langle n^{2} \rangle^{1/2}_{\rm inj}$ (indicated by the colour scheme in the legend). Each downward triangle (left panel) and upward triangle (right panel) corresponds to one of the $100$ random realizations analyzed in Section~\ref{subsec:nrms_and_MC}, for $\dot{K}_{\rm dim} <0$ and $\dot{K}_{\rm dim}>0$, respectively. Stars denote $\langle A_{\rm red}\rangle$ and $\langle \beta \rangle$ for the corresponding $\langle n^{2} \rangle^{1/2}_{\rm inj}$ ensemble of 100 realizations, with the same color scheme for $\langle n^{2} \rangle^{1/2}_{\rm inj}$.}
\label{fig_subsecIII:Recovered_Ramp_Rslope}
\end{figure*}

Fig.~\ref{fig_subsecIII:Recovered_Ramp_Rslope} connects the variance $\langle n^{2} \rangle$ to $A_{\rm red}$ and $\beta$. We plot the pair $(A_{\rm red},\beta)$ measured by \temponest~for each random realization at fixed $\langle n^{2} \rangle^{1/2}_{\rm inj}$ (indicated by the color scheme), with $\dot{K}_{\rm dim} < 0$ (left panel; downward triangles) and $\dot{K}_{\rm dim} > 0$ (right panel; upward triangles). Additionally, we plot the ensemble averages $\langle A_{\rm red} \rangle$ and $\langle \beta \rangle$ as open stars (same colour scheme for $\langle n^{2} \rangle^{1/2}_{\rm inj}$). In both panels, $A_{\rm red}$ and $\beta$ tend to cluster and shift rightward, as $\langle n^{2} \rangle^{1/2}_{\rm inj}$ increases. For instance, in the left panel, $A_{\rm red}$ and $\beta$  span $\approx 4.1$ decades and $2 \leq \beta \leq 10$ for $\langle n^{2} \rangle^{1/2}_{\rm inj}=300.0$, in contrast to $\approx1.4$ decades and $4 \leq \beta \leq 8.3$ for $\langle n^{2} \rangle^{1/2}_{\rm inj}=5985.79$. Similarly, in the right panel, $A_{\rm red}$ and $\beta$ span $\approx 4.2$ decades and $2 \leq \beta \leq 10$, and $\approx 1.8$ decades and $5 \leq \beta \leq 9$, for $\langle n^{2} \rangle^{1/2}_{\rm inj}=300.0$ and $5985.79$, respectively. The results in Fig.~\ref{fig_subsecIII:Recovered_Ramp_Rslope}, consistent with Fig. 4 in \cite{VargasMelatos2023}, imply that one can approximately relate $A_{\rm red}$ and $\beta$ in equation~(\ref{Eq:Temponest_TimingNoise}) to a value of $\langle n^{2} \rangle^{1/2}$. However, given the dispersion in recovered $A_{\rm red}$ and $\beta$ values, there is a limit to how tightly one can tie \temponest~red noise parameters to the dynamical properties of an underlying noise process, be it equations (\ref{Eq:Koft})--(\ref{Eq:finvarianceB}) or something else. A plausible avenue to bypass this limitation is to repeat the procedure described in Section~\ref{subsec:nrms_and_MC} for an observed pulsar, to create a bespoke version of Fig.~\ref{fig_subsecIII:Recovered_Ramp_Rslope}. Then, by comparing the astronomically measured \temponest~noise parameters, $A_{\rm red}$ and $\beta$, with the bespoke version of Fig.~\ref{fig_subsecIII:Recovered_Ramp_Rslope} one can estimate where the astronomical noise realization falls within the $\langle n^{2} \rangle^{1/2}$ range.

The stars in Fig~\ref{fig_subsecIII:Recovered_Ramp_Rslope}, $\langle A_{\rm red} \rangle$ and $\langle \beta \rangle$, show three interesting trends. (i) As  $\langle n^{2} \rangle^{1/2}_{\rm inj}$ decreases, $\langle \beta \rangle$ for $\dot{K}_{\rm dim}<0$ (left panel) decreases, while $\langle \beta \rangle$ for $\dot{K}_{\rm dim}>0$ (right panel) increases. In the left panel, $\langle A_{\rm red}\rangle$ and $\langle \beta \rangle$ span the range $-10.8 \leq \langle A_{\rm red} \rangle / (1~{\rm yr}^{3/2}) \leq -9.49$ and $6.03 \leq \langle \beta \rangle \leq 6.13$. In the right panel, the ranges are $-10.9 \leq \langle A_{\rm red} \rangle / (1~{\rm yr}^{3/2}) \leq -9.46$ and $6.36 \leq \langle \beta \rangle \leq 6.16$. The ranges are consistent with typical values of $A_{\rm red}$ and $\beta$ reported by observational studies, e.g.~\cite{LowerBailes2020} and \cite{ParthasarathyJohnston2020}. (ii) Both panels show $\langle \beta \rangle \sim 6$. This is expected. The Brownian model (\ref{Eq:Koft})--(\ref{Eq:finvarianceB}), in the absence of damping ($\gamma_{\nu}=\gamma_{\dot{\nu}}=\gamma_{\ddot{\nu}}=0$) and secular torques [${\bm E}= 0$ in equation (\ref{Eq:torque_vector})], describes a random walk in $\dot{\nu}$, whose PSD reduces to a power law [functionally identical to $P_{\rm r}(f)$ in (\ref{Eq:Temponest_TimingNoise})] with index $\beta=6$~\citep{Kopeikin1997,AntonelliBasu2022}. (iii) Both panels show that $\langle A_{\rm red} \rangle$ decreases, as $\langle n^{2} \rangle^{1/2}_{\rm inj}$ decreases. That is, on average the amplitude of $P_{\rm r}(f)$ correctly tracks the decrease of the stochastic coefficient $\sigma_{\rm dim}$ (and vice versa). We note that trends (ii) and (iii) are complementary. Both $A_{\rm red}$ and $\beta$ control the overall amplitude of the timing noise process inferred by \temponest.   
%~\citep{BoyntonGroth1972,ShannonCordes2010}
\section{Conclusions}
\label{Sec:conclusions}

In this paper, we study the interplay between two phenomenological modifications of the braking law which can explain anomalous pulsar braking indices. In the first, the proportionality factor in equation~(\ref{Eq:secularpowerlaw}) evolves secularly [scenario (i) in Section~\ref{Sec:Introduction}]. In the second, the secular, power-law braking torque is masked by a stochastic torque, which dominates $\ddot{\nu}$ over typical observational time-scales, viz. equation~(\ref{Eq:ddotnuscreened}) [scenario (ii) in Section~\ref{Sec:Introduction}]. We quantify the interplay between scenarios (i) and (ii) through a combination of analytical calculations, and Monte Carlo simulations involving synthetic data and the Bayesian timing software~\temponest. The synthetic data are generated using a phenomenological secular braking law $\dot{\nu}_{\rm em}(t)=K(t)\nu_{\rm em}(t)^{n_{\rm pl}}$, with $n_{\rm pl}=3$, where the secular evolution of $K(t)$ is described by equation~(\ref{Eq:Koft}). The secular braking is also masked by the stochastic torque $\ddot{\nu}(t)=\ddot{\nu}_{\rm em}(t)+\zeta(t)$, where $\zeta(t)$ is a zero-mean, Langevin driver. The model, fully described by equations~(\ref{Eq:Koft})--(\ref{Eq:finvarianceB}), generates a random realization of synthetic TOAs, given a choice of model parameters $\Xi=\{\sigma^{2}_{\ddot{\nu}},K_{2}/K_{1},\tau_{K}/\tau_{\rm sd}\}$.The synthetic TOAs are analyzed using an initial timing solution from~\tempoDOS~as input for~\temponest. The final timing solution obtained from~\temponest~is converted into a measurement of $n$ for each noise realization of the model and choice of $\Xi$. 

A central result of this paper is equation~(\ref{Eq:theory_nrms}), a predictable, falsifiable formula relating the variance $\langle n^{2} \rangle$ of braking index measurements to a secular coefficient, $\dot{K}_{\rm dim}$ [scenario (i) above; see (\ref{Eq:K_dim})], and a stochastic coefficient, $\sigma_{\rm dim}$ [scenario (ii) above; see (\ref{Eq:Sigma_dim})]. Equation~(\ref{Eq:theory_nrms}) is derived analytically in Appendix~\ref{Appendix:theory} and validated using Monte Carlo simulations involving synthetic data. We use equations~(\ref{Eq:theory_nrms})--(\ref{Eq:Sigma_dim}) to calculate 100 $(\dot{K}_{\rm dim},\sigma_{\rm dim})$ pairs (or equivalently 100 $\Xi$ values) to generate TOA sequences consistent with the injected $\langle n^{2} \rangle^{1/2}_{\rm inj}$. The measured $\langle n^{2} \rangle^{1/2}_{\rm meas}$ obtained using~\temponest~output for the synthetic realizations, agrees well with the injected $\langle n^{2} \rangle^{1/2}_{\rm inj}$. In the anomalous range $3\times10^{2} \leq \langle n^{2} \rangle^{1/2}_{\rm inj} \leq 6\times10^{3}$, we find a mean fractional error between $\langle n^{2} \rangle^{1/2}_{\rm inj}$ an $\langle n^{2} \rangle^{1/2}_{\rm meas}$ of less than $6\%$ averaged over all realizations.  

The agreement between  $\langle n^{2} \rangle^{1/2}_{\rm inj}$ an $\langle n^{2} \rangle^{1/2}_{\rm meas}$ encourages the use of equations~(\ref{Eq:theory_nrms})--(\ref{Eq:Sigma_dim}) to analyze and interpret real astronomical data in the future. For example, one can use equations (\ref{Eq:theory_nrms})--(\ref{Eq:Sigma_dim}) to estimate the physically important quantities $(1-K_2/K_1) \tau_K^{-1}$ and $\sigma_{\ddot{\nu}} \gamma_{\ddot{\nu}}^{-1}$ given an anomalous braking index measurement $|n| \gg 1$ for a real pulsar. One quick approach is to calculate a range of $(\dot{K}_{\rm dim},\sigma_{\rm dim})$ pairs consistent with $n \sim \langle n^{2} \rangle^{1/2}$ from equation~(\ref{Eq:theory_nrms}) and combine the pairs with $T_{\rm obs}$, $\nu_{\rm em}(t_{0})$, $\dot{\nu}_{\rm em}(t_{0})$ and equations (\ref{Eq:K_dim}) and (\ref{Eq:Sigma_dim}) to calculate $(1-K_{2}/K_{1})\tau_{K}^{-1}$ and $\sigma_{\ddot{\nu}}\gamma_{\ddot{\nu}}^{-1}$. Another, more demanding approach is to qualitatively match the astronomical phase residuals with synthetic ones by tuning $\sigma^{2}_{\ddot{\nu}}$, assuming $\gamma_{\ddot{\nu}}=10^{-6}{\rm s}^{-1}$ and $K_{2}=K_{1}$, to estimate  $\langle n^{2} \rangle$ for an assortment of possible $n_{\rm pl}$ values (say $2 \leq n_{\rm pl} \leq 8$, as $n_{\rm pl}$ is not known independently) without assuming $n\sim \langle n^{2} \rangle^{1/2}$. The estimate of $\langle n^{2} \rangle$ is combined with equations (\ref{Eq:K_dim}) and (\ref{Eq:Sigma_dim}) to calculate $(1-K_{2}/K_{1})\tau_{K}^{-1}$ and $\sigma_{\ddot{\nu}}\gamma_{\ddot{\nu}}^{-1}$ per $n_{\rm pl}$ value. Both approaches are complementary. We encourage the reader to select between them to suit the application and resources at hand.

It is always possible to find a set of $(\dot{K}_{\rm dim},\sigma_{\rm dim})$ pairs satisfying equations (\ref{Eq:theory_nrms})--({\ref{Eq:Sigma_dim}}) that are consistent with an astronomical $n$ measurement in a statistical sense, i.e.\ with $n \sim \langle n^2 \rangle^{1/2}$. Importantly, however, some of the pairs produce phase residuals that are inconsistent with the phase residuals measured astronomically. This is a key result of the paper, exemplified by Fig.~\ref{fig_subsecIII:comparison_residuals_Kdim_vs_S} and Fig.~\ref{fig_subsecIII:forced_residuals}: multiple combinations of secular and stochastic anomalies can explain $|n| \gg 1$ as measured in a pulsar in principle, but one must also check whether the phase residuals predicted by each allowed combination match qualitatively the observations. In particular, when emulating PSR J0942$-$5552 in Fig.~\ref{fig_subsecIII:comparison_residuals_Kdim_vs_S} and Fig.~\ref{fig_subsecIII:forced_residuals}, it is found that a meaningful contribution from the stochastic anomaly is necessary to match all the data: scenarios such as $\dot{K}_{\rm dim}=0$ and $\dot{K}_{\rm dim}=\sigma_{\rm dim}$ (and others in between) are consistent with both the $n$ measurement and the phase residuals, whereas the purely secular scenario $\sigma_{\rm dim}=0$ produces phase residuals, which are $\sim 1\%$ of those actually observed, even though it reproduces $n$ correctly.

The PDF of the synthetic $n$ measurements, presented in Section~\ref{subsec:k_sign_and_populations} and Fig.~\ref{fig_subsecIII:hists_kneg_kplus}, reveals three important properties. First, the variance $\langle n^{2} \rangle$ is typically greater than the formal uncertainty $\Delta n$, viz. equation~(\ref{Eq:deltan}), in the anomalous regime $\langle n^{2} \rangle^{1/2}_{\rm inj} \geq 3\times10^{3}$. That is, the blue and cyan histograms in all panels of Fig.~\ref{fig_subsecIII:hists_kneg_kplus} are wider than their corresponding orange histograms. Second, we observe that the number of synthetic measurements which place $n_{\rm pl}$ inside the error bars $n\pm\Delta n$ decreases as $\langle n^{2} \rangle_{\rm inj}^{1/2}$ increases. For example, in the bottom left and top left panels of Fig.~\ref{fig_subsecIII:hists_kneg_kplus}, 15 and 7 realizations out of 100 place $n_{\rm pl}$ within $n\pm\Delta n$. In contrast, in all panels at least 60 out of 100 realizations place $n_{\rm pl}$ within $n\pm\langle n^{2} \rangle^{1/2}_{\rm inj}$. The first and second properties confirm the findings of \cite{VargasMelatos2023}. Third, Fig.~\ref{fig_subsecIII:hists_kneg_kplus} exhibits an asymmetry between the sign of the measured $n$ for $\dot{K}_{\rm dim} < 0$ (left column) and $\dot{K}_{\rm dim} > 0 $ (right column) ensembles. In all the $\dot{K}_{\rm dim} < 0$ panels we find that $\approx 20$ and $\approx 80$ out of 100 values of $n$ are negative and positive, respectively. The opposite is true in all the $\dot{K}_{\rm dim}>0$ panels, where we find $\approx 80$ positive and $\approx 20$ negative values of $n$. The sign asymmetry has been pointed out in observational studies published previously~\citep{ParthasarathyJohnston2020,LowerBailes2020}.

The ideas explored in this methods paper are ready to be tested with real data. However, as foreshadowed above, there are challenges ahead. The experiments performed in Section~\ref{Sec:varandrmsBI} are only possible with synthetic data, where the injected $n_{\rm pl}$ is known. In an astronomical context, $n_{\rm pl}$ is not known independently~\citep{MelroseYuen2016}. There are several ways to bypass this obstacle. One strategy is to do the analysis for a  plausible range of $n_{\rm pl}$, e.g.\ $2 \leq n_{\rm pl} \leq 8$, as described in the third paragraph of this section. Another strategy involves Bayesian model selection on a range of torque models [e.g. only scenario (i), or (ii), or both in Section~\ref{Sec:simulmesnandthemodel}] to determine which $n_{\rm pl}$ is preferred statistically. Some promising steps in this direction include pulsar glitch searches with a hidden Markov model~\citep{MelatosDunn2020,DunnMelatos2022}, and parameter estimation  for a two-component, crust-superfluid model of a neutron star involving a Kalman filter~\citep{MeyersMelatos2021,MeyersO'Neill2021,O'NeillMeyers2024}. All of these techniques, and others currently in development, should be applied to high-quality data sets being generated by the latest pulsar timing campaigns~\citep{NamkhamJaroenjittichai2019a,NamkhamJaroenjittichai2019b,LowerBailes2020,JohnstonSobey2021,ParthasarathyBailes2021}.

\section*{Acknowledgements}

The authors thank Patrick Meyers, for making the {\tt baboo} package freely available, and  Liam Dunn, for guidance in the use of the~\temponest~and~\tempoDOS~software infrastructure. Additionally, we thank Julian Carlin, Liam Dunn, Tom Kimpson, and Joe O'Leary for useful discussions regarding Section~\ref{Sec:varandrmsBI}. This research
was supported by the Australian Research Council Centre of Excellence for Gravitational Wave Discovery (OzGrav), grant number CE170100004. The numerical calculations were performed on the OzSTAR supercomputer facility at Swinburne University of Technology. The OzSTAR program receives funding in part from the Astronomy National Collaborative Research Infrastructure Strategy (NCRIS) allocation provided by the Australian Government.

\section*{Data availability}

The timing solution for PSR J0942$-$5552 comes from \cite{LowerBailes2020}. All the synthetic data are generated using the open access software package {\tt baboo} available at \url{http://www.github.com/meyers-academic/baboo}~\citep{MeyersO'Neill2021}. We use~\tempoDOS~\citep{HobbsEdwards2006} and~\temponest~\citep{LentatiAlexander2014} to analyze the synthetic data.

%%%%%%%%%%%%%%%%%%%%%%%%%%%%%%%%%%%%%%%%%%%%%%%%%%%%%%%%%%%%%%%%%%%%%%%%%%%%%%%%%%%%%%%%%%%%%%

\twocolumn

\bibliographystyle{mnras}
\bibliography{ADSABS_bib, main_non_ads_bib}

%%%%%%%%%%%%%%%%%%%%%%%%%%%%%%%%%%%%%%%%%%%%%%%%%%

%%%%%%%%%%%%%%%%% APPENDICES %%%%%%%%%%%%%%%%%%%%%
%\newpage
\appendix

\section{Theory of anomalous braking indices due to secular and stochastic torques}
\label{Appendix:theory}

In this appendix, we present an analytic theory of anomalous braking indices, which describes the statistics of the $n$ values when measured in the presence of secular and stochastic torques. The central result is a predictive, falsifiable formula relating the variance $\langle n^{2} \rangle$ of braking index measurements [see (\ref{Eq:theory_nrms})] to a secular-torque term involving $K_{\rm dim}$ [see (\ref{Eq:K_dim})] and a stochastic-torque term involving $\sigma_{\rm dim}$ [see (\ref{Eq:Sigma_dim})], generalizing the predictive formula with zero secular evolution presented in Equation (14) in~\cite{VargasMelatos2023}. To derive the formula, we present in Appendix~\ref{Appendix:Analyticalsol} the analytic solution for the Brownian model defined by equations (\ref{Eq:Koft})--(\ref{Eq:finvarianceB}). For a thorough justification of the Brownian model we refer the reader to Appendix A1 of~\cite{VargasMelatos2023}. In Appendix~\ref{Appendix:variance_n} we combine the analytic solutions of (\ref{Eq:Koft})--(\ref{Eq:finvarianceB}) with a standard `non-local' measurement of $n$, based on a finite difference formula involving $\dot{\nu}(t_{0})$ and $\dot{\nu}(t_{0}+T_{\rm obs})$, to rigorously derive an expression for the variance $\langle n^{2} \rangle$.

\subsection{Analytic solution of the Brownian model (\ref{Eq:Koft})--(\ref{Eq:finvarianceB})}
\label{Appendix:Analyticalsol}

The mathematical form of the Brownian model treated in this paper is based on the one described by~\cite{VargasMelatos2023}. The difference is that $K(t)$ in equation (\ref{Eq:secularpowerlaw}) depends on time in this paper, as specified in equation (\ref{Eq:Koft}), whereas $K(t)$ is constant in~\cite{VargasMelatos2023}. As explained in Appendix A1 of \cite{VargasMelatos2023}, the stochastic equation of motion~(\ref{Eq:setofequations}) can be solved after multiplying by an exponential integrating factor, subject to the initial conditions $\nu(t_{0})=\nu_{\rm em}(t_{0})$, $\dot{\nu}(t_{0})=\dot{\nu}_{\rm em}(t_{0})$, and $\ddot{\nu}(t_{0})=\ddot{\nu}_{\rm em}(t_{0})$, and $t_{0}=0$. The result is

\begin{align}
    \phi(t) &= \phi(0)+\int_{0}^{t} dt'\,\nu(t')\label{Eq_Apndx1:phi} \\
    \nu(t) &= \nu_{\rm em}(t)+e^{-\gamma_{\nu}t}\int_{0}^{t} dt' e^{\gamma_{\nu} t'} \left[ \dot{\nu}(t')-\dot{\nu}_{\rm em}(t')\right],\label{Eq:appnu} \\
    \dot{\nu}(t) &= \dot{\nu}_{\rm em}(t)+e^{-\gamma_{\dot{\nu}}t}\int_{0}^{t} dt' e^{\gamma_{\dot{\nu}} t'} \left[ \ddot{\nu}(t')-\ddot{\nu}_{\rm em}(t')\right],\label{Eq:appdot_nu} \\
    \ddot{\nu}(t) &= \ddot{\nu}_{\rm em}(t)+e^{-\gamma_{\ddot{\nu}}t} \int_{0}^{t} dt' e^{\gamma_{\ddot{\nu}}t'} \xi(t')\label{Eq:appddot_nu}.
\end{align}

We take the initial phase to be $\phi(0)=0$ without loss of generality.

The secular braking evolution is calculated by solving equations~(\ref{Eq:secularpowerlaw}) and (\ref{Eq:Koft}) for $\nu_{\rm em}(t)$, viz.

\begin{equation}
    \nu_{\rm em}(t) = \nu_{\rm em}(0)f(t)^{(1-n_{\rm pl})^{-1}}, \label{Eq:appnuSEC}
\end{equation}

with 

\begin{equation}
    f(t) = 1+\frac{K_{2}}{K_{1}}\frac{t}{\tau_{\rm sd}}+\frac{\tau_{K}}{\tau_{\rm sd}}\Big[1-\frac{K(t)}{K_{1}}\Big],
   %f(t) = 1+\frac{1}{\tau_{\rm sd}}\Bigg[ \Bigg(\frac{K_{2}}{K_{1}} \Bigg)t+\tau_{K}\Bigg(1-\frac{K_{2}}{K_{1}} \Bigg)\Bigg(1-e^{-t/\tau_{K}}\Bigg) \Bigg],
    \label{Eq:appffunc}
\end{equation}

where the characteristic spin down age is given by

\begin{equation}
    \tau_{\rm sd}=-\frac{\nu_{\rm em}(0)}{(n_{\rm pl}-1)\dot{\nu}_{\rm em}(0)}.
    \label{Eq:apptau_sd}
\end{equation}

%The evolution of the rotational parameters due the secular torque is complete by calculating the first, second, and third time-derivatives of $\nu_{\rm em}(t)$, namely
By differentiating equation (\ref{Eq:appnuSEC}) once, twice, and thrice with respect to $t$, we obtain explicit expressions for the secular terms in equations (\ref{Eq:appnu})--(\ref{Eq:appddot_nu}), viz.

\begin{align}
    \dot{\nu}_{\rm em}(t) =& \dot{\nu}_{\rm em}(0)\frac{K(t)}{K_{1}}f(t)^{\frac{n_{\rm pl}}{1-n_{\rm pl}}}, \label{Eq:appdotnuSEC} \\
    \ddot{\nu}_{\rm em}(t) =& \ddot{\nu}_{0}\Bigg\{\frac{K(t)^{2}}{K_{1}^{2}}+\frac{\tau_{\rm sd}}{\tau_{K}}\frac{(n_{\rm pl}-1)}{n_{\rm pl} K_{1}}\Big[ K(t)-K_{2}\Big]f(t)\Bigg\} f(t)^{\frac{2n_{\rm pl}-1}{1-n_{\rm pl}}},\label{Eq:appddotnuSEC} \\
    \dddot{\nu}_{\rm em}(t) =& \dddot{\nu}_{0} \Bigg\{\frac{K(t)}{K_{1}}\Bigg[\frac{K(t)^{2}}{K_{1}^{2}}+\frac{\tau_{\rm sd}}{\tau_{K}}\frac{(n_{\rm pl}-1)}{n_{\rm pl}K_{1}}\Big(K(t)-K_{2}\Big)f(t)\Bigg]  \nonumber \\
    &-\frac{n_{\rm pl}-1}{n_{\rm pl}(2n_{\rm pl}-1)K_{1}}\Big(K(t)-K_{2}\Big)\Bigg[(n_{\rm pl}+1)\frac{\tau_{\rm sd}}{\tau_{K}}\frac{K(t)}{K_{1}} \nonumber \\
    &+(n_{\rm pl}-1)\frac{\tau_{\rm sd}^{2}}{\tau_{K}^{2}}f(t)\Bigg]f(t)\Bigg\}f(t)^{\frac{3n_{\rm pl}-2}{1-n_{\rm pl}}}. \label{Eq:appdddotnuSEC}
\end{align}

In equations~(\ref{Eq:appddotnuSEC})--(\ref{Eq:appdddotnuSEC}), the parameters $\ddot{\nu}_{0}$ and $\dddot{\nu}_{0}$ are given by $\ddot{\nu}_{0}=n_{\rm pl}\dot{\nu}_{\rm em}(0)^{2}/\nu_{\rm em}(0)$ and $\dddot{\nu}_{0}=n_{\rm pl}(2n_{\rm pl}-1)\dot{\nu}_{\rm em}(0)^{3}/\nu_{\rm em}(0)^{2}$. 

With the secular torque included, we now have $\ddot{\nu}_{\rm em}(0) \neq \ddot{\nu}_{0} $ and $ \dddot{\nu}_{\rm em}(0) \neq \dddot{\nu}_{0}$, with 

\begin{equation}
   \ddot{\nu}_{\rm em}(0) = \ddot{\nu}_{0}\Bigg[ 1+\frac{\tau_{\rm sd}}{\tau_{K}} \frac{(n_{\rm pl}-1)}{n_{\rm pl}} \Bigg(1-\frac{K_{2}}{K_{1}} \Bigg) \Bigg], 
   \label{Eq:app_initial_ddotnu}
\end{equation}

and 

\begin{align}
    \dddot{\nu}_{\rm em}(0) =& \dddot{\nu}_{0}\Bigg\{1+\frac{(n_{\rm pl}-1)}{n_{\rm pl}(2n_{\rm pl}-1)}\Bigg(1-\frac{K_{2}}{K_{1}} \Bigg) \nonumber \\
    &\times\Bigg[(n_{\rm pl}-2)\frac{\tau_{\rm sd}}{\tau_{K}}-(n_{\rm pl}-1)\frac{\tau_{\rm sd}^{2}}{\tau_{K}^{2}} \Bigg] \Bigg\}. 
    \label{Eq:app_initial_dddotnu}
\end{align}

Equations~(\ref{Eq:app_initial_ddotnu}) and (\ref{Eq:app_initial_dddotnu}) lead to two important consequences in the regime where one has $\tau_{\rm sd}/\tau_{K} \gg 1$ and there is no timing noise ($\sigma_{\rm dim}=0$). First, the initial conditions $\ddot{\nu}_{\rm em}(0)$ and $\dddot{\nu}_{\rm em}(0)$ satisfy $\vert \ddot{\nu}_{\rm em}(0) \vert \gg \vert \ddot{\nu}_{0} \vert$, and $\vert \dddot{\nu}_{\rm em}(0) \vert \gg \vert \dddot{\nu}_{0} \vert$. As explained in Sections~\ref{Sec:simulmesnandthemodel} and~\ref{Sec:varandrmsBI}, the former condition on $\ddot{\nu}_{\rm em}(0)$ generates anomalous braking indices with $n=\nu_{\rm em}(0)\ddot{\nu}_{\rm em}(0)/\dot{\nu}_{\rm em}(0) \gg n_{\rm pl}$, a longstanding result in the existing literature~\citep{Goldreich1970,BlandfordRomani1988,LinkEpstein1997,Melatos2000,TaurisKonar2001,PonsVigano2012,JohnstonKarastergiou2017,WassermanCordes2022}. Second, the sign of $1-K_{2}/K_{1}$ defines the sign of $n$ through $\ddot{\nu}_{\rm em}(0)$. When astrophysically motivated stochastic torques are included ($\sigma_{\rm dim} \gg 1$) the sign of $\ddot{\nu}(t)$ (and hence the measured $n$) differs in general from the sign of $\ddot{\nu}_{\rm em}(t)$. For $K_{2}/K_{1} > 1$ ($\dot{K}_{\rm dim} < 0$) the measured $n$ are predominantly negative, while the opposite is true for $K_{2}/K_{1} < 1$ ($\dot{K}_{\rm dim} > 0$) (see Section~\ref{subsec:k_sign_and_populations}).

The zero-mean fluctuating variables $\delta \nu(t)=\nu(t)-\nu_{\rm em}(t)$, $\delta \dot{\nu}(t)= \dot{\nu}(t)-\dot{\nu}_{\rm em}(t)$, and $\delta \ddot{\nu}(t)=  \ddot{\nu}(t)-\ddot{\nu}_{\rm em}(t)$ in equations (\ref{Eq:appnu})--(\ref{Eq:appddot_nu}) obey the same solution as in~\cite{VargasMelatos2023}. Hence the covariances $\langle \delta \dot{\nu}(t)^{2} \rangle, \langle \delta \dot{\nu}(t) \ddot{\nu}(t) \rangle$, and $\langle \delta \ddot{\nu}(t)^{2} \rangle$ are also the same as in the latter reference. Specifically, in the regime where the fluctuation amplitudes match the observed timing behaviour of typical pulsars [as exemplified by Figs.~\ref{fig_subsecII:example_f2_walk}(b),~\ref{fig_subsecIII:comparison_residuals_Kdim_vs_S}, and~\ref{fig_subsecIII:forced_residuals}] defined by $\gamma_{\nu}\sim\gamma_{\dot{\nu}}\ll T^{-1}_{\rm obs} \ll \gamma_{\ddot{\nu}}$, we find 

\begin{align}
    \langle \delta \ddot{\nu}(t)^{2} \rangle &=  \frac{\sigma^{2}_{\ddot{\nu}}}{2\gamma_{\ddot{\nu}}}, \label{Eq_Apndx1:var_ddot_nu_inf}\\
    \langle \delta \dot{\nu}(t)^{2} \rangle &=  \frac{\sigma^{2}_{\ddot{\nu}}}{2\gamma_{\dot{\nu}}\gamma_{\ddot{\nu}}^{2}}\left(1-e^{-2\gamma_{\dot{\nu}}t}\right), \label{Eq_Apndx1:var_dot_nu_inf}\\ 
    \langle \delta \dot{\nu}(t) \ddot{\nu}(t) \rangle &=  \frac{\sigma^{2}_{\ddot{\nu}}}{2\gamma_{\ddot{\nu}}^{2}} \label{Eq_Apndx1:covar_ddot_nu_dot_nu_inf}.
\end{align}

Equations (\ref{Eq_Apndx1:var_ddot_nu_inf})--(\ref{Eq_Apndx1:covar_ddot_nu_dot_nu_inf}) follow from equations (A15)--(A17) in \cite{VargasMelatos2023}.

\subsection{Variance $\langle n^{2} \rangle$}
\label{Appendix:variance_n}

 An astronomical measurement of $n$ corresponds to a specific, time-ordered, random realization of the stochastic torque in a neutron star, which corresponds in turn to a random realization of the Brownian increment $d{\bf B}(t)$ in equation (\ref{Eq:setofequations}) in the theory. There is no way to know where the actual realization in the neutron star lies within the statistical ensemble from which it is drawn. Therefore, when using the theory to make falsifiable predictions, one must calculate PDFs of the relevant observables or, more compactly, their ensemble averages. Testing the predictions involves checking for statistical consistency, e.g.\ whether or not a measured $n$ value falls within the probable range $|n - \langle n  \rangle | \lesssim \langle n^2 \rangle^{1/2}$. 
 
 In this paper, we focus on the variance $\langle n^2 \rangle$. We also focus on $n$ measurements conducted via the `non-local' approach favored by previous authors~\citep{JohnstonGalloway1999,VargasMelatos2023}. The distinction between `non-local' and `local' measurements of $n$ is important and subtle. It is discussed thoroughly in Appendix A of \cite{VargasMelatos2023}.

In the non-local measurement of $n$, $\ddot{\nu}(t)$ is evaluated by finite diferencing the time series $\dot{\nu}(t)$, at times $t_{1}$ and $t_{2}=t_{1}+T_{\rm obs}$, as opposed to considering the instantaneous second derivative $X_{4}(t)=\ddot{\nu}(t)$ which contains unresolvable random fluctuations on arbitrary short time-scales due to the Langevin term $\xi(t)$ in Equation~(\ref{Eq:appdot_nu})~[see Section~\ref{subsec:Themodel} and Appendix A1 in~\cite{VargasMelatos2023}]. From equation~(6) in \cite{JohnstonGalloway1999}, we have

\begin{equation}
    n=1-\frac{\dot{\nu}(t_{1})\nu(t_{2})-\dot{\nu}(t_{2})\nu(t_{1})}{\dot{\nu}(t_{1})\dot{\nu}(t_{2})T_{\rm obs}}.
    \label{Eq:app_n_nonloc}
\end{equation}

In the regime $\gamma_{\nu}\sim\gamma_{\dot{\nu}}\ll T^{-1}_{\rm obs} \ll \gamma_{\ddot{\nu}}$, which corresponds to the astrophysically observed fluctuations regime $\vert \delta \nu(t) \vert \ll \vert \nu_{\rm em}(t) \vert$ and $\vert \delta \dot{\nu}(t) \vert \ll \vert \dot{\nu}_{\rm em}(t) \vert$, one can write equation~(\ref{Eq:app_n_nonloc}) as

\begin{align}
    n =&\,1 -\frac{1}{T_{\rm obs}} \frac{\nu_{\rm em}(t_{2})}{\dot{\nu}_{\rm em}(t_{2})} \left[1+\frac{\delta \nu(t_{2})}{\nu_{\rm em}(t_{2})}-\frac{\delta \dot{\nu}(t_{2})}{\dot{\nu}_{\rm em}(t_{2})} \right] \nonumber \\
    &+\frac{1}{T_{\rm obs}}\frac{\nu_{\rm em}(t_{1})}{\dot{\nu}_{\rm em}(t_{1})} \left[1+\frac{\delta \nu(t_{1})}{\nu_{\rm em}(t_{1})}-\frac{\delta \dot{\nu}(t_{1})}{\dot{\nu}_{\rm em}(t_{1})} \right]. \label{Eq:appn_expanded_flucs_1_2}
\end{align}

The ensemble average of equation~(\ref{Eq:appn_expanded_flucs_1_2}) yields 

\begin{equation}
    \langle n \rangle = 1 - \frac{1}{T_{\rm obs}}\Bigg[\frac{\nu_{\rm em}(t_{2})}{\dot{\nu}_{\rm em}(t_{2})}-\frac{\nu_{\rm em}(t_{1})}{\dot{\nu}_{\rm em}(t_{1})} \Bigg], \label{Eq:app_mean_n_no_approxSEC}
\end{equation}

which can be simplified by noticing that one has

\begin{equation}
    \frac{\nu(t_{2})K(t_{2})}{\dot{\nu}(t_{2})}-\frac{\nu(t_{1})K(t_{1})}{\dot{\nu}(t_{1})} = (1-n_{\rm pl})\int_{t_{1}}^{t_{2}} dt'K(t')
    \label{Eq:appaux_Koft}
\end{equation}

by integrating equation~(\ref{Eq:secularpowerlaw}). The time-scale $\tau_{K}$ for the secular evolution of $K(t)$ satisfies $T_{\rm obs} \ll \tau_{K} \ll \tau_{\rm sd}$, so a Taylor expansion of $K(t_{2})$ around $t_{2}=t_{1}$ in equation (\ref{Eq:appaux_Koft}) yields

\begin{align}
    \langle n \rangle =& 1-\frac{1}{K(t_{1})}\Bigg\{\Bigg[-\dot{K}(t_{1})-\frac{1}{2}\ddot{K}(t_{1})T_{\rm obs} \Bigg]\frac{\nu_{\rm em}(t_{2})}{\dot{\nu}_{\rm em}(t_{2})} \nonumber \\
    &+(1-n_{\rm pl})\Bigg[K(t_{1})+\frac{1}{2}\dot{K}(t_{1})T_{\rm obs}+\frac{1}{6}\ddot{K}(t_{1})T_{\rm obs}^{2}\Bigg]\Bigg\}. \label{Eq:app_n_approx_taylor_2ndorder}
    %&= n_{\rm pl}+\frac{\dot{K}(t_{1})}{K(t_{1})}\frac{\nu(t_{2})}{\dot{\nu}(t_{2})}+{\cal O}(T_{\rm obs}/\tau_{K}) \nonumber
\end{align}

To leading order, equation~(\ref{Eq:app_n_approx_taylor_2ndorder}) reduces to

\begin{align}
    \langle n \rangle =& n_{\rm pl}+\frac{\dot{K}(t_{1})}{K(t_{1})}\frac{\nu_{\rm em}(t_{2})}{\dot{\nu}_{\rm em}(t_{2})} \label{Eq:app_mean_nSEC} \\
    \approx& n_{\rm pl}+\frac{\dot{K}(t_{1})}{K(t_{1})}\frac{\nu_{\rm em}(t_{1})}{\dot{\nu}_{\rm em}(t_{1})} \label{eq:app_sec_coeff}.
    % }. 
\end{align}

Equations (\ref{Eq:app_mean_nSEC}) and (\ref{eq:app_sec_coeff}) are equal up to corrections of order ${\cal O}(T_{\rm obs}/\tau_{\rm sd}) \ll 1$. The rightmost term in equation (\ref{eq:app_sec_coeff}) reduces to $\dot{K}_{\rm dim}$ in equation (\ref{Eq:K_dim}) for the specific secular evolution described by (\ref{Eq:Koft}) with~(\ref{Eq:apptau_sd}). Equations~(\ref{Eq:app_mean_n_no_approxSEC})--(\ref{eq:app_sec_coeff}) do not assume a specific functional form for $K(t)$, e.g. equation~(\ref{Eq:Koft}). Any other functional form which satisfies $\vert \dot{K}(t)/K(t) \vert \sim \tau^{-1}_{K}$ leads to equation~(\ref{eq:app_sec_coeff}).

Upon combining equations~(\ref{Eq:appn_expanded_flucs_1_2}) and ~(\ref{Eq:app_mean_nSEC}), we obtain an expression for $n-\langle n \rangle$. The expression can be simplified by noting that $\vert \delta \nu(t) / \nu_{\rm em}(t) \vert \ll \vert \delta \dot{\nu}(t) / \dot{\nu}_{\rm em}(t) \vert$ holds empirically for all pulsars observed to date~\citep{LowerBailes2020,ParthasarathyJohnston2020} (see also Fig.~\ref{fig_subsecII:example_f2_walk}), and we have $\nu(t_{1})/\nu(t_{2})=1+{\cal O}(T_{\rm obs}/\tau_{\rm sd}) \approx 1$. Applying these inequalities, we combine equations~(\ref{Eq:appn_expanded_flucs_1_2}) and (\ref{eq:app_sec_coeff}) to obtain

\begin{equation}
    n-\langle n \rangle \approx \frac{\nu_{\rm em}(t_{1})}{\dot{\nu}_{\rm em}(t_{1})^{2}}\left[ \frac{\delta \dot{\nu}(t_{2})-\delta \dot{\nu}(t_{1})}{T_{\rm obs}} \right].
    \label{Eq:app_n_minus_nSEC}
\end{equation}

In order to calculate $\langle n^{2} \rangle$, we square equation~(\ref{Eq:app_n_minus_nSEC}) and calculate $\langle [ \delta \dot{\nu}(t_{2})-\delta \dot{\nu}(t_{1}) ]^{2} \rangle$ by setting $t=T_{\rm obs}$ in equation~(\ref{Eq_Apndx1:var_dot_nu_inf}), with the time origin arbitrary (say $t_{1}=0$). In the astrophysically relevant regime $\gamma_{\dot{\nu}}T_{\rm obs} \ll 1$, the above steps combined with equation~(\ref{Eq:app_n_minus_nSEC}) yield

\begin{equation}
    \langle n^{2} \rangle - \langle n \rangle^{2} = \frac{\sigma^{2}_{\ddot{\nu}}\nu_{\rm em}(0)^{2}}{\gamma^{2}_{\ddot{\nu}}\dot{\nu}_{\rm em}(0)^{4}T_{\rm obs}}.\label{Eq:app_var_n}
\end{equation}

The right-hand side of equation~(\ref{Eq:app_var_n}) is the square of $\sigma_{\rm dim}$ as defined in equation~(\ref{Eq:Sigma_dim}).

Combining equations~(\ref{eq:app_sec_coeff}) and (\ref{Eq:app_var_n}), we write

\begin{equation}
    \langle n^{2} \rangle = \Big(n_{\rm pl}+\dot{K}_{\rm dim}\Big)^{2}+\sigma_{\rm dim},
    \label{Eq:apptheory_nrms}
\end{equation}

which is identical to equation~(\ref{Eq:theory_nrms}).  It reduces to equation (14) in \cite{VargasMelatos2023} in the special case $\dot{K}_{\rm dim} = 0$. The variance $\langle n^{2} \rangle$ in equation~(\ref{Eq:apptheory_nrms}) is closely related to the fractional dispersion across the ensemble, ${\rm DISP}(n)$, given by equation~(13) in~\cite{VargasMelatos2023}. 

If the proportionality constant $K(t)$ changes with time ($\dot{K}_{\rm dim} \neq 0$) equation~(\ref{Eq:apptheory_nrms}) predicts that the standard deviation of the measured braking indices is $\sigma_{\rm dim}$, while the mean is $n_{\rm pl}+\dot{K}_{\rm dim}$. Hence the anomalous regime $\vert n \vert \gg 1 $ occurs for $\dot{K}_{\rm dim} \gg 1, \sigma_{\rm dim} \gg 1$, or both.

% Don't change these lines
\bsp	% typesetting comment
\label{lastpage}
\end{document}